\documentclass[prd,aps,nofootinbib,notitlepage]{revtex4}
\usepackage{graphicx,epsf,amsmath,amsfonts,amssymb,amsbsy,textcomp}
\usepackage{epsfig}
\usepackage{epstopdf}
\newcommand{\vev}[1]{\langle#1\rangle}
\newcommand{\mat}{\left ( \begin{array}}
\newcommand{\emat}{\end{array} \right )}
\newcommand{\vect}{\left ( \begin{array}{c}}
\newcommand{\evect}{\end{array} \right )}

\begin{document}

\title{%Dense baryonic matter and dualities of QCD phase diagram (with applications)%Applications of 
Dense Baryonic Matter and Applications of QCD % please define if appropriate
Phase~Diagram Dualities %of QCD % please define if appropriate
%Phase~Diagram
}

% Author Orchid ID: enter ID or remove command
%\newcommand{\orcidauthorA}{0000-0000-000-000X} % Add \orcidA{} behind the author's name
%\newcommand{\orcidauthorB}{0000-0000-000-000X} % Add \orcidB{} behind the author's name

% Authors, for~the paper (add full first names)
\author{Tamaz G. Khunjua $^{1,2}$, {Konstantin  G.} Klimenko $^{3}$ and {Roman N.} Zhokhov $^{3, 4}$}
%Please carefully check the accuracy of aALL the names and affiliations.

% Authors, for~metadata in PDF
%\AuthorNames{T. G. Khunjua, K. G. Klimenko  and R. N. Zhokhov}

%\address{%
\affiliation{
$^{1}$ The University of Georgia, GE-0171 Tbilisi, Georgia; gtamaz@gmail.com}
%Please add the specific department/school/faculty/campus.
\affiliation{$^{2}$ Department of Theoretical Physics, A. Razmadze Mathematical Institute, I. Javakhishvili Tbilisi State University, GE-0177 Tbilisi, Georgia}
\affiliation{$^{3}$ Logunov Institute for High Energy Physics, NRC ``Kurchatov Institute'', 142281 Protvino, Russia; Konstantin.Klimenko@ihep.ru}
%Please add post code. 
%Is the Moscow Region/Moscow extra?
%$^{2}$ \quad Faculty of Physics, M.V. Lomonosov Moscow State University, Moscow\\
\affiliation{$^{4}$  Pushkov Institute of Terrestrial Magnetism, Ionosphere and Radiowave Propagation (IZMIRAN), Troitsk, 142190 Moscow, Russia, zhokhovr@gmail.com}   %Please add post code. %Is the Moscow Region/Moscow extra?

% Contact information of the corresponding author
%\corres{Correspondence: zhokhovr@gmail.com; Tel.: +7-9035041060}

% Current address and/or shared authorship
%\firstnote{Current address: Affiliation 3} 
%\secondnote{These authors contributed equally to this work.}
% The commands \thirdnote{} till \eighthnote{} are available for further notes

%\simplesumm{} % Simple summary

%\conference{} % An extended version of a conference paper

% Abstract (Do not insert blank lines, i.e. \\) 
\begin{abstract}%
%Recently it has been found that there is a duality of QCD phase diagram it was revealed first in the QCD motivated toy model are discussed 
Recently it has been found that quantum chromodynamics (QCD) phase diagram possesses a duality between chiral symmetry breaking and pion condensation. For the first time this was revealed in the QCD motivated toy model. Then it was demonstrated in effective models as well and new additional dualities being found. We briefly recap the main features of this story and then discuss its applications as a tool to explore the QCD phase structure. The~most appealing application is the possibility of getting the results on the QCD phase diagram at large baryon density. Taking the idea from large $1/N_{c}$ universalities it was argued that the scenario of circumventing the sign problem with the help of dualities seems %feasible 
plausible. It~is also discussed that there is a %long standing
persistent problem about whether there should be catalysis or anti-catalysis of chiral symmetry breaking by chiral imbalance. One can probably say that the issue is settled after lattice results (first principle approach%, though performed at unphysically large pion mass
), where the catalysis was observed. But they used an unphysically large pion mass so  it is still interesting to get additional indications that this is the case. It~is shown just by the duality property that there exists catalysis of chiral symmetry breaking. So, having in mind our results and the earlier lattice simulations, one can probably claim that this issue is settled.  It is demonstrated that the duality can be used to obtain new results. As an example, it~is showcased how the phase structure of dense quark matter with chiral imbalance (with possibility of inhomogeneous phases) can be obtained from the  knowledge of a QCD phase diagram with isopin~asymmetry.
%First, it~is briefly overviewed the similar to our dualities obtained in the so-called large $1/N_{c}$ orbifold equivalence principle. And then the possibility of circumventing the sign problem has been expanded to our dualities and it is argued that it is a feasible scenario. It~is well known that the domain of QCD phase diagram where there is large baryon density is notoriously hard to be predicted theoretically due to the so called sign problem. And first principle investigations is up to now unavailable in this region. So to get hints of the phase structure of QCD in this respect one can, for~example, have use of effective models. There are discussed the recently found dualities of QCD phase diagram.  There are discussions that this problem can be partially solved or alleviated by the interesting properties of the phase structure, namely dualities. But the dualities can be used in other way and can have a number of applications. Here we discuss the dualities of QCD phase diagram and the and their possible applications to getting hints about phase structure.  
\end{abstract}

% Keywords
%\keyword{QCD phase diagram, non-zero baryon density, chiral imbalance, dualities}

% The fields PACS, MSC, and JEL may be left empty or commented out if not applicable
%\PACS{J0101}
%\MSC{}
%\JEL{}

\maketitle

\section{Introduction}\label{sec1}

It is believed now that the dynamics of mesons and baryons should be described by the quantum chromodynamics (QCD), non-Abelian gauge theory of quarks and gluons. The~ultimate goal of QCD studies is to understand all the hadronic phenomena at the level of quarks and gluons. For very high energy processes, where the perturbation theory works, this paradigm has been very successful, for~example, describing the deep inelastic scattering (DIS) of leptons and hadrons.
The situation is different at low energies (of the order of
low lying hadron masses (1 GeV)), where one needs non-perturbative description and analytic study is very complicated.
Although QCD has been a central point of the high energy physics community from its inception, it remains one of the main topics today. Not only because it is still impossible to truly understand the mechanism of confinement and non-perturbative physics but also because of the increasing interest %of the twenty thirty years 
in the studies of QCD phase structure in extreme conditions%at finite temperature and density
.
%In 1973, with the discovery of asymptotic freedom, QCD was at the frontier of particle physics studies. Now, in 2006, QCD is strongly installed (returning) at the center of particle physics researches. This, not only because of the abundance of data on jet emission at HERA and Tevatron, but

One can easily think about at least two important external parameters for QCD in equilibrium, namely temperature $T$ and baryon chemical potential $\mu_{B}$ (conjugated to baryon density $n_B$). Now~let us try to guess the orders of magnitude of temperature and density %that could lead to interesting implications
that could cause interesting phenomena. Recall now that the intrinsic energy scale of QCD is $\Lambda_{QCD}\sim 200$ MeV (the energy at which coupling constant blows up). So one can expect that the phase transition of thermal QCD should take place around the temperature $T\sim \Lambda_{QCD}$ around 200 MeV (that is not that far from lattice simulation results)
 and a phase transition of dense QCD should happen at a baryon density of the order of $n_{B}\sim \Lambda_{QCD}^3\sim 1$ fm$^{-3}$.

Finite baryon density QCD has piqued huge interest in nuclear physics, high energy physics and even astrophysics. It~is excepted that QCD has a very rich phase structure in (T, $\mu$) parameter space \cite{Mannarelli, Ayala:2011vs, Ayala1, Inagaki, Fujihara:2008ae, Fujihara:2008wx, Friesen:2014mha, Nedelko:2014sla, Blaschke, Shahrbaf:2019vtf,  Bauswein:2019skm, Alvarez-Castillo:2018rrv, Radzhabov:2010dd, Tawfik:2019tkp, Rajagopal:1999cp,  Sasaki:2009zh, Grigorian:2017xqd, Kolomeitsev:2017gli}, 
 %Please check the references order, and Ref. 51 isn't mentioned in the text.
 and there are several heavy-ion collision experiments such as NICA, FAIR, RHIC (BES~II), J-PARC, HIAF % please define the abbreviations
 that will elucidate the properties of dense QCD matter %picture
in the near future. Especially awaited is the NICA (Nuclotron-based Ion Collider fAility) complex that is now under construction at  the Joint Institute for Nuclear Research (Dubna, Russia) \cite{NICAWhitePaper}.

But it is not hard to believe that temperature and baryon density are not the only relevant parameters in various physical setups.
These physical settings, for~example, could be systems with a large isospin imbalance \cite{Mannarelli, KogutSinclair, BrandtEndrodi, BrandtEndrodi1}. Consider, for~instance, the~initial state of heavy-ion collisions, the~initial ions have twice as many neutrons as protons and this can be important in collisions of not that high energy. Moreover,  neutron star matter is characterized by even larger isospin imbalance (it~consists of mostly neutrons and the fraction of protons is rather small%with rather small proton fraction
). More unexpected is the fact that, although it is believed that baryon density in the early Universe is typically small, a large lepton asymmetry (poorly constrained by observations) might lead to a large
isospin imbalance \cite{Schwarz:2009ii}.

There is another new interesting field of novel transport phenomena, so-called  anomalous transport phenomena, that has caused a lot of excitement in the community. Heavy-ion collision experiments have exhibited intriguing hints of possible signals but due to background effects the situation is still unclear. 
One of the central players in this field is the so-called chiral imbalance $n_{5}$ (the~difference between densities of 
right-handed and left-handed quarks) or  
the corresponding chiral chemical potential $\mu_{5}$. It~is believed that the chiral imbalance can be created at high temperatures in the heavy-ion collisions due to the Adler-Bell-Jackiw 
anomaly and nontrivial gluon field configurations. Moreover, in strong magnetic field or under rotation, due to the so-called chiral separation \cite{Metlitski} or chiral vortical \cite{Fukushima:2018grm} effects, chiral density can be produced in dense quark matter \cite{Khunjua:2018jmn, Khunjua:2018dbm, KhunjuaMoscowUnivBull}. Chiral imbalance can also be generated in parallel electric and magnetic fields \cite{Ruggieri:2016fny, RuggieriPeng, Ruggieri:2016fny1}.
Moreover, there is another type of chiral imbalance, when chiral 
densities of $u$ and $d$ quarks are different, it~is called chiral isospin imbalance $n_{I5}$. It~can be shown that chiral isospin imbalance can be produced in magnetized or rotating dense quark matter \cite{Khunjua:2018jmn}. Moreover, in parallel electric and magnetic fields chiral isospin imbalance can be generated as well. 
Let us also note that in the context of a QCD phase diagram, the formal inclusion of chiral isospin imbalance is more rigorous than the chiral one.
There are a lot of studies on chiral imbalanced QCD~\cite{Braguta,  Braguta13, Braguta1, Braguta:2016aov, andrianov, GattoRuggieri, Liu, Liu4, RuggieriandPeng, Zhuang, Suenaga:2019jqu}.
%There's no citation for 17 between them, please check.

In this letter 
we discuss the dualities of QCD phase structure and its possible uses and applications to the process of unraveling the puzzles of QCD phase diagram including the region of large baryon densities.
First, the~duality between chiral symmetry breaking (CSB) and charged pion condensation (PC) phenomena is considered in terms of QCD related toy models, namely the NJL$_2$ model, where it has been found for the first time. Then it is shown that duality holds in the framework of effective model for QCD that bolster our confidence that it can be valid in real QCD. It~is also shown that the duality remains valid if one considers the phase diagram with all four possible imbalances, baryon density, isospin, chiral isospin  and  chiral imbalance (first, it~was shown only for the first three of them). It~can be considered as another indication that it is not a coincidence and there is something behind it. After~that it is argued in the framework of effective model that there exist other dualities of phase structure not as strong as the main one but also rather interesting.

Then comes the main part of the paper; the discussion on how and where the dualities can be used and be helpful in understanding the phase structure of QCD.% at large baryon density

\begin{enumerate}
\item[(i)]   First, there is a brief overview of dualities similar to ours (universalities) obtained in the so-called large $1/N_{c}$ orbifold equivalence principle. Then the idea of the possibility of circumventing the sign problem has been expanded to our dualities and it is argued that it is a feasible scenario.
\item[(ii)]   It is shown that a problem of catalysis or anti-catalysis of chiral symmetry breaking by chiral imbalance can be resolved just by duality and the rather well-established knowledge of pion condensation properties at isospin density.
\item[(iii)]   It is shown that the duality can be used to produce new results and new phase diagrams (different sections of the phase diagram). As an example, it~is showcased how, from the phase structure of dense quark matter with non-zero isospin density (including the possibility of inhomogeneous  condensates (phases)), one can obtain, based on the duality only, the~phase structure of dense quark matter with chiral imbalance.
\end{enumerate}

In this way, from the different regions studied in a number of works, whether one can assemble the whole picture of the phase structure of QCD at finite baryon and isospin density including inhomogeneous phases was explored.

%%%%%%%%%%%%%%%%%%%%%%%%%%%%%%%%%

Most of the studies in this paper are performed in the framework of the effective NJL model. It~is one of the most widely used effective models for QCD, where one can explore the phase structure at non-zero baryon density. %In the one can study different phenomena, influence of magnetic field on the chiral phase transition, different phenomena.
A lot of different phenomena have been successfully studied in this approach.
Despite all that, the NJL model has a number of limitations and drawbacks, there are no gluons in the consideration and it is not confining, but this can be partially improved by including interaction with constant background gluon field and considering the Polyakov loop as an order parameter for deconfinement (the so-called PNJL model). %sin
Also, %As a rule 
often the considerations are performed in the mean field approximation (ignoring the meson fluctuations). In this mean field approach one struggles to explain various phenomena. For example, the~behaviour of chiral condensate %depends 
is flat at its origin (it weakly depends on temperature). This behavior is drastically different from what is obtained from ChPT %please define if appropriate
 \cite{GasserLeutwyler}.
%Please revise the order of the references citations. All the references should be cited in the numerical order in the text. Please check the citation order throughout the text.
One can obtain the right low-temperature parabolic behaviour of chiral condensate %melting
only if one includes into consideration the meson loops (go beyond the mean field) \cite{Florkowski:1996wf}. In the simplest version of the NJL model in the mean field %it is also 
one also gets the wrong prediction for the influence of magnetic field on the chiral phase transition. It~is shown on the lattice that there should be an inverse magnetic catalysis (IMC) effect  \cite{Bali:2011qj}, %Please revise the order of the references citations. All the references should be cited in the numerical order in the text. Please check the citation order throughout the text.
that is, it is found from all the observables that the
pseudo-critical temperature decreases significantly with the increase of magnetic field. Mean field NJL consideration predicts the magnetic catalysis (MC) (increase of pseudo-critical temperature). The correct behaviour can be obtained only if one phenomenologically extends the model to include the dependence of the NJL model interaction coupling on the magnetic field \cite{Endrodi, Ferreira, Ferreira:2014kpa} (it should decrease with the magnetic field and mimic the expected running of the coupling with the strength of the magnetic
field). %The MC effect on the light quark condensates is obtained for low and high temperatures and the IMC effect appears near the transition temperature region 
In this approach, one can obtain MC at low and high temperatures and IMC around the critical temperature. %One can also obtain 
These results can also be obtained in the NJL model beyond mean field approximation \cite{Mao:2016fha}, where meson contribution lead to the dependence of effective coupling on both the magnetic field and the temperature.
Having all these remarks in mind one can see that, despite all the successes of the mean field NJL model, %predictions
it has many limitations and one should always remember this.

%e reproduced by introducing a magnetic field dependence directly on the scalar interaction coupling of the NJL/PNJL models. The~decreasing of Gs(eB) with the magnetic field is essential, within effective quark models, to mimic the expected running of the coupling with the magnetic field strength.

%We show that the IMC is obtained if the interaction strength between quarks decreases with the magnetic field. We propose two mechanisms that reproduce the IMC effect, which assume a weakening of the scalar coupling with increasing magnetic field strength.
%%%%%%%%%%%%%%%%%%%%%%%%%%%%%%%%%%

Let us briefly overview the content of the paper and what is covered in each section.
Section~\ref{sec1} outlines the introduction and the purpose of this paper.
In Section~\ref{sec2} the Gross-Neveu model and its different extensions, including the NJL$_2$ model, are discussed.  Section~\ref{sec3} contains the discussion of quark matter with non-zero baryon density, isospin, chiral and chiral isospin imbalances and corresponding charges.
The phase structure and its duality in the framework of the (1+1)-dimensional QCD related toy model are discussed in Section~\ref{sec3.1}. Then, in Section~\ref{sec3.2} duality is shown to be valid in the framework of the effective model. Section~\ref{sec3.3} contains the proof that the duality remains valid even if we include chiral imbalance (in addition to chiral isospin one) in the system.  In Section~\ref{sec3.4} it is shown that there are other dualities. In Section~\ref{sec4} it is demonstrated how duality can be used and how it can help us study the phase structure of QCD at finite densities.

\section{(1+1)-Dimensional Models: The GN Model and Its Extensions%NJL$_{2}$ model
}\label{sec2}
In order to understand the phase structure of matter at finite temperature and baryon density, it~is necessary to comprehend the non-perturbative vacuum of QCD and its properties.
As has been pointed out, QCD is hard to deal with, which is why one can try to study the phase structure in a similar but simpler and hence tractable model (QCD related toy models).

%The Gross-Neveu model [1] resembles QCD in many respects and can be solved analytically in the limit of an inﬁnite number of fermion flavours N.

\subsection{GN Model}\label{sec2.1}
The Gross-Neveu (GN) model %[12, 48, 49, 60]
is a model with  four-fermion interaction that consists of only a single quark flavour \cite{Gross:1974jv, Winstel:2019zfn, Feinberg}. It~is remarkable that it can be solved analytically in the limit of an infinite number of quark colours $N_{c}$. Interest in this model from the particle physics side stems from the fact that it in many respects resembles QCD%, theory of strong interaction describing (interacting via gluons) quarks
. For example, it~exhibits a lot of similar inherent features to QCD such as renormalizability, asymptotic freedom, dynamical chiral symmetry breaking (in vacuum) and its restoration (at finite temperatures), dimensional transmutation, and meson and baryon bound states \cite{Schnetz:2005ih}.
In addition, the~$\mu_B-T$ phase diagram is 
qualitatively the same.

Probably an even more unexpected fact is that (1+1)-dimensional GN type models have exhibited great success in the description of a variety of quasi-one-dimensional condensed matter systems~\cite{caldas,Schnetz:2005ih,Thies:2005wv,MertschingFischbeck, Machida}, %, ranging from the Peierls-Fr¨ohlich model [4] 
for~example, polyacetylene \cite{caldas,Schnetz:2005ih} or similar models can be used in the description of planar systems~\cite{pl, Klimenko:2012qi,Klimenko:2012tk, Klimenko:2013gua},
%Finally, GN models with ﬁnite N have been found useful for describing electrons in 
carbon nanotubes and fullerenes \cite{Lin, Kalinkin, kolmakov}.

%\cite{Thies:2005wv}

 %The Lagrangian of the GN model in Euclidean space is
% Let us further mention that (1+1)-dimensional Gross-Neveu type models are also suitable for the description of physics in quasi one-dimensional condensed matter systems like polyacetylene \cite{caldas,Schnetz:2005ih}

%Interest from the particle physics side stems from the fact that the simple Lagrangian (1) shares nontrivial properties with quantum chromodynamics (QCD), notably asymptotic freedom, dimensional transmutation, meson and baryon bound states, chiral symmetry breaking in the vacuum as well as its restoration at high temperature and density (for a pedagogical review, see [2]).
%Perhaps even more surprising and less widely appreciated is the fact that GN type models have enjoyed considerable success in describing a variety of quasi-onedimensional condensed matter systems, ranging from the Peierls-Fr¨ohlich model [4] over ferromagnetic superconductors [5] to conducting polymers, e.g. doped trans-polyacetylene [6]. By way of example, the~kink and kink-antikink baryons ﬁrst derived in ﬁeld theory in the chiral limit (m0 = 0) [7] have been important for understanding the role of solitons and polarons in electrical conductivity properties of doped polymers. Likewise, one can show that baryons in the massive GN model (m0 6= 0) are closely related to polarons and bipolarons in polymers with non-degenerate ground states, e.g. cis-polyacetylene [8, 9, 10]. Finally, GN models with ﬁnite N have been found useful for describing electrons in carbon nanotubes and fullerenes [11, 12]. 

The Lagrangian of the GN model has the form
\begin{eqnarray}
&&  L=\mathrm{i}\bar q\gamma^\nu\partial_\nu q + \frac
{G}{N_c}(\bar qq)^2 ,  \label{1}
\end{eqnarray}
where the quark field $q(x)\equiv q_{i\alpha}(x)$ is a colour $N_c$-plet ($\alpha=1,...,N_c$) as
well as a two-component Dirac spinor (the summation in (\ref{1})
over color, and spinor indices is implied). The~Dirac $\gamma^\nu$-matrices ($\nu=0,1$) and $\gamma^5$ in (1) are matrices in 
two-dimensional spinor space,
\begin{equation}
\begin{split}
\gamma^0=
\begin{pmatrix}
0&1\\
1&0\\
\end{pmatrix};\qquad
\gamma^1=
\begin{pmatrix}
0&-1\\
1&0\\
\end{pmatrix};\qquad
\gamma^5=\gamma^0\gamma^1=
\begin{pmatrix}
1&0\\
0&{-1}\\
\end{pmatrix}.
\end{split}
\end{equation}
\subsection{Chiral GN Model ($\chi$GN)}\label{sec2.2}
A straightforward extension of the GN model (\ref{1}) can be obtained by adding to the Lagrangian a pseudo-scalar term \cite{Thies:2003zr, Basar:2009fg}. It~is called a chiral GN ($\chi$GN) model) and its Lagrangian would take the following form
\begin{eqnarray}
&&  L=\mathrm{i}\bar q\gamma^\nu\partial_\nu q+ \frac
{G}{N_c}\Big [(\bar qq)^2+(\mathrm{i}\bar q\gamma^5 q)^2 \Big
], 
\end{eqnarray}
%This model contains a scalar ﬁeld combination ¯ ψjψj, which corresponds to a σ-like particle, and a pseudoscalar ﬁeld combination ¯ ψjıγ5ψj, which corresponds to an η-like particle. 
It can be shown that it is invariant under continuous chiral symmetry transformations $U_A(1)$: $\psi\to e^{i\gamma^{5}\theta} \psi$. There are also certain similarities between this model and one-flavour QCD, if one excludes the chiral anomaly from consideration.

\subsection{NJL$_2$ Model}\label{sec2.3}
Let us even further generalize the GN model by considering, in addition to scalar channel, pion- like field combinations \cite{Thies:2019ejd,kkzz, %kkzzp,
Khunjua:2019ini, Ebert:2016hkd, Khunjua:2018vyb}.  The Lagrangian of this so-called (1+1)-dimensional Nambu-Jona-Lasinio model (NJL$_2$ model) is
\begin{eqnarray}
&&  L=\mathrm{i}\bar q\gamma^\nu\partial_\nu q+ \frac
{G}{N_c}\Big ((\bar qq)^2+(\bar q\mathrm{i}\gamma^5\vec\tau q)^2 \Big
),
\end{eqnarray}
In addition to all the similarities with QCD that were discussed for the GN model, this model can describe the interactions of pions and is similar to two-flavour QCD.
We consider this model in order to mimic the phase structure of real dense quark matter with two massless quark flavors ($u$ and $d$ quarks). Below~the phase diagram of dense quark matter with isospin and chiral isospin imbalance will be studied in the framework of this model.

\section{Dense Quark Matter with Isospin and Chiral Imbalance}\label{sec3}
If one wants to describe quark matter with non-zero baryon (quark) density one needs to add to the Lagrangian the following term $\frac{\mu_B}{3}\bar q\gamma^0q$. %This leads to the settings to describe the matter with non-zero difference between number of baryons (quarks) and anti-baryons (anti-quarks). 
So $\mu_B$ is baryon chemical potential which leads to the settings to describe the matter with a non-zero difference between number of baryons and anti-baryons. Sometimes different quantities and quark chemical potential $\mu=\frac{\mu_B}{3}$ are used, which describe the non-zero difference between number of quarks and anti-quarks.
If in addition, one has the isospin imbalance in the system, that is, a different number of protons and neutrons (or equivalently u quarks and d quarks) one needs to add to the Lagrangian the following term $\frac{\mu_I}2\bar q \tau_3\gamma^0q$. The~more exotic opportunity that will be considered in the following is chiral imbalance, the~difference between left-handed and right-handed quarks in the system. %For this
In order to describe it one needs to add to the Lagrangian the following term $\mu_5\bar q\gamma^{0}\gamma^{5}q$. There still another imbalance that will be considered below, namely chiral isospin imbalance  $\mu_{I5}$, that accounts for the difference between chiral imbalances of different flavours ($u$ and $d$ quarks), $n_{I5}=n_{5}^{u}-n_{5}^{d}\neq0$ and it is introduced as the following term $\frac{\mu_{I5}}2\bar q \tau_3\gamma^0\gamma^5q$ in the Lagrangian.

\subsection{Dense Isospin Asymmetric Quark Matter with Non-Zero Chirality: Phase Diagram in QCD Related Model}\label{sec3.1}
Bearing in mind all that has been said in the above two sections
now let us consider the phase diagram of dense ($\mu_{B}\neq0$) quark matter with non-zero isospin ($\mu_{I}\neq0$) and chiral isospin ($\mu_{I5}\neq0$) imbalance in the framework of the (1+1)-dimensional QCD related toy model, namely the NJL$_2$ model. 
Let us also stress that here the chiral imbalance $\mu_{5}$ of the system is considered to be zero, so the Lagrangian in this case has the following form
\begin{eqnarray}
&&  L=\bar q\Big [\gamma^\nu\mathrm{i}\partial_\nu
+\frac{\mu_B}{3}\gamma^0+\frac{\mu_I}2 \tau_3\gamma^0+\frac{\mu_{I5}}2 \tau_3\gamma^0\gamma^5\Big ]q+ \frac
{G}{N_c}\Big [(\bar qq)^2+(\bar q\mathrm{i}\gamma^5\vec\tau q)^2 \Big
].  \label{01}
\end{eqnarray}
As said above, all these parameters ($\mu_B,\mu_I,\mu_{I5}$) are introduced in order to investigate in the framework of the model quark matter with nonzero baryon $n_B$, isospin $n_I$ and axial isospin $n_{I5}$ densities, respectively.
 
It is evident that at zero $\mu_B=\mu_I=\mu_{I5}=0$ the Lagrangian is invariant with respect to $SU(2)_{L}\times SU(2)_{R}\times U_B(1)$ group.
If one introduce  non-zero $\mu_B\neq0$ into the system the symmetries remain the same. But if we include non-zero isospin imbalance $\mu_I\neq0$ then the symmetry of the model is
$U_B(1)\times U_{I_3}(1)\times U_{AI_3}(1)$, where $U_{I_3}(1):~q\to\exp (\mathrm{i}\beta\tau_3/2) q$ and $U_{AI_3}(1):~q\to\exp (\mathrm{i}
\omega\gamma^5\tau_3/2) q$.

If in addition, one introduces non-zero $\mu_{I5}$ then the symmetry group will not change.

 So the quark bilinears $\frac 13\bar q\gamma^0q$, $\frac 12\bar q\gamma^0\tau^3 q$ and $\frac 12\bar q\gamma^0\gamma^5\tau^3 q$ are the zero components of corresponding to these groups conserved currents and their ground state expectation values are just the baryon, isospin and chiral isospin densities, that is, $n_B=\frac 13\vev{\bar q\gamma^0q}$, $n_I=\frac 12\vev{\bar q\gamma^0\tau^3 q}$
and $n_{I5}=\frac 12\vev{\bar q\gamma^0\gamma^5\tau^3 q}$. As~usual, the~quantities $n_B$, $n_I$ and $n_{I5}$ can also be found by differentiating the thermodynamic potential of the system with respect to the corresponding chemical potentials. For brevity we will use the following notations $\mu\equiv\mu_B/3$, $\nu\equiv\mu_I/2$ and $\nu_{5}\equiv\mu_{I5}/2$.

In order to find the thermodynamic potential of the system, it~is more convenient to use a semi-bosonized version of the Lagrangian, which contains composite bosonic fields $\sigma (x)$ and $\pi_a (x)$ $(a=1,2,3)$ 
\begin{eqnarray}
\widetilde{L} &=& \bar q\Big [\gamma^\rho\mathrm{i}\partial_\rho
+\mu\gamma^0 + \nu\tau_3\gamma^0+\nu_{5}\tau_3\gamma^0\gamma^5-\sigma
-\mathrm{i}\gamma^5\pi_a\tau_a\Big ]q
 -\frac{N_c}{4G}\Big [\sigma\sigma+\pi_a\pi_a\Big ].
\label{2}
\end{eqnarray}
From the Lagrangian (\ref{2}) one can get the Euler--Lagrange equations of the bosonic fields
\begin{eqnarray}
\sigma(x)=-2\frac G{N_c}(\bar qq);~~~\pi_a (x)=-2\frac G{N_c}(\bar q
\mathrm{i}\gamma^5\tau_a q).
\label{200}
\end{eqnarray}
The composite bosonic field $\pi_3 (x)$ can be identified
with the physical $\pi_0$ meson, whereas the $\pi^\pm
(x)$- meson fields with the following combinations of the composite fields, 
%%%%%%% I remember opposite signs for $\pi^\pm$ :  \pi_1 (x) -/+ i\pi_2 (x). Check!
$\pi^\pm (x)=(\pi_1 (x)\mp i\pi_2 (x))/\sqrt{2}$. 
%%%%%%%%%%%%%%%%%%%%%%%%%%%%%%%%%%%%%%%%%%%%%%%%%%%%%%%%%%%%%%%
In general, the~phase structure is characterized by the behaviour of so-called order parameters (or condensates) with respect to external parameters such as temperature, chemical potentials, and so forth. In our case such order parameters are the ground state expectation values of the composite fields, $\vev{\sigma (x)}$ and $\vev{\pi_a (x)}$ $(a=1,2,3)$.

If $M=\vev{\sigma(x)}\ne 0$ (or $\vev{\pi_3(x)}\ne 0$), then the axial isospin $U_{AI_3}(1)$ symmetry (remnant of chiral symmetry at non-zero $\mu_{I}$ and $\mu_{I5}$) is dynamically broken down and
$$
U_B(1)\times U_{I_3}(1)\times U_{AI_3}(1)\to U_B(1)\times U_{I_3}(1).
$$

Whereas if $\Delta=\vev{\pi_1(x)}\ne 0$ (or $\vev{\pi_2(x)}\ne 0$) we have a spontaneous breaking of the isospin symmetry $U_{I_3}(1)$ and
$$
U_B(1)\times U_{I_3}(1)\times U_{AI_3}(1)\to U_B(1)\times U_{AI_3}(1).
$$
 Since in this case condensates of the fields $\pi^+(x)$ and $\pi^-(x)$ are not zero, this phase is usually called the charged pion condensation (PC) phase.

Starting from the linearized semi-bosonized model Lagrangian (\ref{2}), one can obtain in the leading order
of the large $N_c$-expansion (i.e., in the one-fermion loop
approximation) the thermodynamic
potential (TDP) $\Omega (M,\Delta%\sigma,\pi_a
)$ of the system:
\begin{eqnarray}
\Omega (M, \Delta%\sigma,\pi_a
)\equiv -\frac{{\cal S}_{\rm {eff}}(\sigma,\pi_a)}{N_c\int
d^2x}\bigg |_{~\{\sigma,~\pi_1,~\pi_2,~\pi_3\}=\{M,~ \Delta,~ 0,~0\}%\rm {const}
}
=\frac{M^2+\Delta^2}{4G}+%\mathrm{i}{\rm Tr}_{sf}\int\frac{d^2p}{(2\pi)^2}\ln\overline{D}(p),
\mathrm{i}\int\frac{d^2p}{(2\pi)^2}\ln
P_4(p_0),
\label{7}
\end{eqnarray}
where $P_4(p_0)=%\epsilon_1\epsilon_2\epsilon_3\epsilon_4=
\eta^4-2a\eta^2-b\eta+c$, $\eta=p_0+\mu$ and
\begin{eqnarray}
a&=&M^2+\Delta^2+p_1^2+\nu^2+\nu_{5}^2;~~b=8p_1\nu\nu_{5};\nonumber\\
c&=&a^2-4p_1^2(\nu^2+\nu_5^2)-4M^2\nu^2-4\Delta^2\nu_5^2-4\nu^2\nu_5^2.
\label{10}
\end{eqnarray}

One can see that the TDP %please define
 is invariant with respect to the so-called duality transformation
\begin{eqnarray}
{\cal D}:~~~~M\longleftrightarrow \Delta,~~\nu\longleftrightarrow\nu_5.
 \label{16}
\end{eqnarray}
The duality tells us that we can simultaneously exchange chiral condensate and charged pion condensate and isospin and chiral imbalances and the results do not change. This means that chiral symmetry breaking phenomenon in the system with isospin (chiral) imbalance is the same as (equivalent to) charged pion condensation phenomenon in the system with chiral (isospin) imbalance. It~is an interesting property of the phase structure of the toy model and possibly of real QCD.

It is clear from (\ref{7}) that the effective potential is an ultraviolet (UV) divergent, so we need to renormalize it. This procedure can be found in Reference \cite{kkzz,  %kkzzp,
Khunjua:2019ini, Ebert:2016hkd, Khunjua:2018vyb} and it is shown that it does not concern the duality property, which can be seen already at the level of unrenormalized TDP. The~phase structure of the model is considered in detail in Reference \cite{kkzz, %kkzzp,
Ebert:2016hkd, Khunjua:2018vyb}.

\subsection{Dense Isospin Asymmetric Quark Matter with Non-Zero Chirality: Effective Model Consideration}\label{sec3.2}
In the previous section it was shown in the framework of the NJL$_{2}$ model that there is a duality between chiral symmetry breaking and pion condensation phenomena.
Although the NJL$_{2}$ model has a lot of features in common with QCD and %probably the duality is one of them and one can try to argue that the duality can be a property of QCD
 one can probably argue that the duality is one of them, the NJL$_{2}$ model is not real QCD and one cannot guarantee that all the properties that it has can be transformed to QCD. So it is interesting to try to check whether there is a duality in effective models for QCD. At least they are (3+1) dimensional and they have many more connections with QCD.

Here in this section
%Then 
the same situation, that is, the quark matter with isospin and chiral isospin imbalance, has been considered in the framework of a more realistic effective model for QCD, namely the Nambu--Jona-Lasinio model. Its Lagrangian has the similar form
\begin{eqnarray}
&&  L=\bar q\Big [\gamma^\nu\mathrm{i}\partial_\nu
+\frac{\mu_B}{3}\gamma^0+\frac{\mu_I}2 \tau_3\gamma^0+\frac{\mu_{I5}}2 \tau_3\gamma^0\gamma^5\Big ]q+ \frac
{G}{N_c}\Big [(\bar qq)^2+(\bar q\mathrm{i}\gamma^5\vec\tau q)^2 \Big
],  \label{14}
\end{eqnarray}
where, in contrast to the (1+1)-dimensional case, the flavor doublet, $q=(q_u,q_d)^T$, ($q_u$ and $q_d$ u and d quark fields) are four-component Dirac spinors (also color $N_c$-plets) and the gamma matrices are normal, familiar (3+1)-dimensional ones.

One can also use the semi-bosonized Lagrangian and the technique similar to that which was used above and to obtain the TDP of the model in this case. %But it is simpler to consider
After a rather long but straightforward calculations one can show that in this case the TDP of the model reads
\begin{eqnarray}
\Omega (M,\Delta)~=\frac{M^2+\Delta^2}{4G}+\mathrm{i}\int\frac{d^4p}{(2\pi)^4}P_-(p_0)P_+(p_0),
\end{eqnarray}
where
$P_-(p_0)P_+(p_0)\equiv\big (\eta^4-2a\eta^2-b\eta+c\big )\big (\eta^4-2a\eta^2+b\eta+c\big )$ and $\eta=p_0+\mu$, $|\vec p|=\sqrt{p_1^2+p_2^2+p_3^2}$.
%We also used the following notations
We also used the same notations for $a,b$ and $c$ just with a substitution $p_{1}\to|\vec p|$
\begin{eqnarray}
a&&=M^2+\Delta^2+|\vec p|^2+\nu^2+\nu_{5}^2;~~b=8|\vec p|\nu\nu_{5};\nonumber\\
c&&=a^2-4|\vec p|^2(\nu^2+\nu_5^2)-4M^2\nu^2-4\Delta^2\nu_5^2-4\nu^2\nu_5^2.
\label{10}
\end{eqnarray}
One can see in a similar way that the duality (\ref{16}) takes place in this case as well, meaning it can be found even in the more realistic effective model for QCD. So one can conclude %this shows 
that the duality is probably the property of the phase structure of real QCD.

The phase structure of the model (\ref{14}) %in this case 
is discussed in \cite{Symmetry}.

\subsection{Inclusion of $\mu_{5}$ Chiral Imbalance and the Consideration of the General Case}\label{sec3.3}
The duality property of the phase structure of quark matter was shown in the case of non-zero baryon density and isospin imbalance of the system as well as chiral isospin imbalance. But it is not obvious in any sense that this property holds in other situations and is universal. For example, there~are a lot of studies (see Reference \cite{Fukushima:2008xe} and references therein) discussing the possibility of generation of chiral imbalance $\mu_{5}$ (in reality much more than the discussions on the generation of $\mu_{I5}$) and it is interesting to investigate whether the duality also holds for the general case of the phase structure of quark matter with all four chemical potentials (baryon density and three imbalances—isospin, chiral and chiral isospin). If~the duality is valid for the phase structure of this general system then it is probably more deep quality of the phase structure.

%This consideration has been actually performed in \cite{}. It~has been performed in the framework of effective model for QCD, NJL model, wich Lagrangian has the from
Let us try to consider this situation in the framework of an effective model for QCD (NJL model),  which Lagrangian has the form
\begin{eqnarray}
&&  L=\bar q\Big [\gamma^\nu\mathrm{i}\partial_\nu
+\frac{\mu_B}{3}\gamma^0+\frac{\mu_I}2 \tau_3\gamma^0+\mu_{5} \gamma^0\gamma^5+\frac{\mu_{I5}}2 \tau_3\gamma^0\gamma^5\Big ]q+ \frac
{G}{N_c}\Big [(\bar qq)^2+(\bar q\mathrm{i}\gamma^5\vec\tau q)^2 \Big
].
\end{eqnarray}
It is almost the same Lagrangian as (\ref{1}) but containing the chiral chemical potential $\mu_{5}$ accounting for the chiral imbalance $n_{5}$.

%In \cite{} there has been
One can obtain the TDP of the system in this rather general case and it was shown to have the~form
\begin{eqnarray}
\Omega (M,\Delta)~=\frac{M^2+\Delta^2}{4G}+\mathrm{i}\int\frac{d^4p}{(2\pi)^4}P_-(p_0)P_+(p_0),
\label{TDP}
\end{eqnarray}
where
$P_+(\eta)P_-(\eta)\equiv\big (\eta^4-2a_+\eta^2+b_+\eta+c_+\big )\big (\eta^4-2a_-\eta^2+b_-\eta+c_-\big )$ and
\begin{eqnarray}
a_\pm&&=M^2+\Delta^2+(|\vec p|\pm\mu_{5})^2+\nu^2+\nu_{5}^2;~~b_\pm=\pm 8(|\vec p|\pm\mu_{5})\nu\nu_{5};\nonumber\\
c_\pm&&=a_\pm^2-4 \nu ^2
\left(M^2+(|\vec p|\pm\mu_{5})^2\right)-4 \nu_{5}^2 \left(\Delta ^2+(|\vec p|\pm\mu_{5})^2\right)-4\nu^{2} \nu_{5}^2.
\label{TDP10}
\end{eqnarray}
One can see that the duality property ${\cal D}:~~~M\longleftrightarrow \Delta,~~\nu\longleftrightarrow\nu_5~$ stays the same in this case of non-zero chiral imbalance (non-zero $\mu_{5}\neq0$) as well. So the duality is the property of the phase structure of dense quark matter with isospin, chiral and chiral isospin imbalances (in the rather general case one can imagine) at least in terms of effective model.

\subsection{Other Dualities}\label{sec3.4}
One can also note that there are more dualities aside from the one we have already mentioned. Although they are not as strong as the original one described above they are still pretty interesting and useful.
These new dualities %They
are valid only if there are some additional constraints, for~example, there is no pion condensation in the system or chiral symmetry is restored.
One should show additionally that these constraints are fulfilled dynamically in considered situation, for example, chiral symmetry is dynamically restored (high temperature or density).

Let us discuss it in more detail.
One can show from Equations (\ref{TDP}) and (\ref{TDP10}) that at the constraint $\Delta=0$ (if there is no charged pion condensation in the system)
\begingroup\makeatletter\def\f@size{9.2}\check@mathfonts
\def\maketag@@@#1{\hbox{\m@th\normalsize\normalfont#1}}%
\begin{eqnarray}
&& P_+(\eta)P_-(\eta)\Big |_{\Delta=0}=\big [M^2+(|\vec p|+\mu_5+\nu_5)^2-(\eta+\nu)^2\big ]
\big [M^2+(|\vec p|+\mu_5-\nu_5)^2-
(\eta-\nu)^2\big ]\times\nonumber\\
&&
\times\big [M^2+(|\vec p|-\mu_5+\nu_5)^2-(\eta-\nu)^2\big ]\big [M^2+(|\vec p|-\mu_5-\nu_5)^2-(\eta+\nu)^2\big ]
\label{18}
\end{eqnarray}
\endgroup
and one can see that the TDP (\ref{TDP}) in this case %at this constraint 
is invariant with respect to the following transformation
\begin{eqnarray}
{\cal D}_M:~~~~\Delta=0,~~~\mu_5\longleftrightarrow\nu_{5}.
 \label{19}
\end{eqnarray}
This is another duality and it shows that chiral symmetry breaking phenomenon does not feel the difference between two types of chiral imbalance (chiral and chiral isospin imbalances).

Likewise, it~is possible to demonstrate that the integrand in the expression for the TDP (\ref{TDP}) with the constraint $M=0$ has the following form
\begingroup\makeatletter\def\f@size{9.2}\check@mathfonts
\def\maketag@@@#1{\hbox{\m@th\normalsize\normalfont#1}}%
\begin{eqnarray}
&& P_+(\eta)P_-(\eta)\Big |_{M=0}=\big [\Delta^2+(|\vec p|+\mu_5+\nu)^2-(\eta+\nu_5)^2\big ]
\big [\Delta^2+(|\vec p|+\mu_5-\nu)^2-
(\eta-\nu_5)^2\big ]\times\nonumber\\
&&
\times\big [\Delta^2+(|\vec p|-\mu_5+\nu)^2-(\eta-\nu_5)^2\big ]\big [\Delta^2+(|\vec p|-\mu_5-\nu)^2-(\eta+\nu_5)^2\big ].
\label{20}
\end{eqnarray}
\endgroup
So at the constraint $M=0$ the TDP is invariant under the transformation
\begin{eqnarray}
{\cal D}_\Delta:~~~~M=0,~~~\mu_5\longleftrightarrow\nu.
 \label{21}
\end{eqnarray}
That shows that charged pion condensation phenomenon is influenced in the same way by chiral and isospin imbalance. Systems with isospin and chiral imbalance are completely different systems and it is remarkable that some phenomena in these systems are entirely equivalent.

Additionally, one can notice that the dualities ${\cal D}_M$ and ${\cal D}_\Delta$ are themselves dual to each other with respect to the ${\cal D}$ duality (\ref{16}). So one can conclude that there exists only one independent additional duality and one can get the other duality by making use of the main duality (\ref{16}).
%%%%%%%%%%%%%%%%%%%%%%%%%%%%%%%%%%%%%%%
%The diagram clarifying the dualities and their interrelations is depicted in Figure 1.

\section{Use of Dualities}\label{sec4}
\vspace{-6pt}
\subsection{Circumventing the Sign Problem with Use of Dualities}\label{sec4.1}
%Orbifold \cite{Cherman:2010jj, Cherman:2011mh}

One of the key open questions in the standard model of particle physics is the phase structure of QCD, especially at non-zero baryon density. However, the~understanding of the properties of QCD phase diagram at finite baryon chemical potential is limited by the so-called sign problem. It~consists of the fact that at $\mu_{B}\neq0$ the fermion determinant is no longer real-valued and positive quantity and the conventional Monte-Carlo simulations are impossible in this case. There are a number of approaches to solve or at least %ameliorate
alleviate the sign problem  but it stands %till
to this day in the way of obtaining the phase structure of dense matter from first principles. And it is quite likely that it will not be solved completely in the near future.%it will stay for some time

Here we will discuss a possibility to circumvent it in a way and possibly get some clues of phase structure at large baryon densities.
It has been noticed that there is a whole class of gauge theories that have no sign problem even at nonzero baryon chemical potential and probably may resemble QCD ($SU(3)$ gauge theory). The~examples include %include QCD with isospin chemical potential µI [1], two-color QCD with even degenerate
%flavors Nf at µB 6= 0 [2, 3], 
%two-colour QCD with even
$SU(3)$ theory but with fermions in the adjoint representation, $SO(2N_c)$
gauge theory and $Sp(2N_c)$ gauge theory or two-colour QCD with even number of flavours
$N_{f}$ (with the same mass). So one can study the properties of these theories not encountering the sign problem. But it
is not clear how, if at all, these theories are connected with real QCD.
Then~it has been shown in References \cite{Cherman:2010jj, increase, Cherman:2011mh} by means of the orbifold equivalence technique that in the limit of large $N_c$ (large number of colours) the whole or at least part of the phase diagram of these theories are the same. This sameness of the phase structure is called universality. There exist pieces of evidence that this universality remains valid even for QCD with three-colour but only approximately~\cite{Cherman:2010jj,  Cherman:2011mh}. These~universalities are very similar to our dualities but they connect not only phase structure at different chemical potentials but also the phase structure of different gauge theories. If the gauge groups of two theories related by the universality are the same then they coincide with our dualities. %For example, one of the universalities, namely the equivalence of phase structure of QCD at finite $\nu_{B}$ and at finite $\mu_{I}$ outside the BEC-BCS crossover region is very similar with our duality (\ref{19}).
For example, one of the universalities, namely the equivalence of phase structure of QCD at finite $\mu_{5}$ and at finite $\mu_{I}$ where the pion condensate and chiral condensate should be exchanged is very similar to the duality that can be obtained from our dualities for these cases.

It was also pointed out in References \cite{Cherman:2010jj, Cherman:2011mh} that the universalities of phase structure can be used in circumventing the sign problem due to the fact that one gauge theory can be sign problem free. For example, universality can relate gauge theories with groups $G_{1}$ and  $G_{2}$ at different chemical potentials $\mu_{1}$ and $\mu_{2}$, then assume that $G_{1}=SU(3)$ and %this theory is QCD 
and $\mu_{1}=\mu_{B}$, so it is QCD at finite baryon density. If $G_{2}$ happens to be sign problem free (at non-zero $\mu_{2}$) theory then one can obtain the phase structure of QCD at $\mu_{B}$ by studying the $G_{2}$ gauge theory at $\mu_{2}$ using lattice simulations. The~same idea can be applied to our dualities. For example,  QCD at isospin  chemical potential $\mu_{I}$ or QCD at chiral chemical potential $\mu_{5}$ has no sign problem. Phase structures of QCD at different  chemical potentials are connected with each other by dualities and there are ideas that the dualities can concern also baryon chemical potential $\mu_{B}$ (connect $\mu_{B}$ with other chemical potentials). Here we only note that it is a viable option and the detailed discussion is left for the future. 

Let us make another small remark here at the end of the section. As we have pointed out above there are hints that the universalities are also valid approximately for the case $N_{c}=3$. Let us note that some arguments (although, maybe not that strong) can be made from NJL model considerations.
If the same (to the corresponding large $N_{c}$ equivalence) duality can be obtained in NJL model then one can conclude that it holds not only for large $N_{c}$ limit but in mean field approximation as well. In~mean field approximation one can take $N_{c}=3$ (let us put aside the argument that mean field is a good approximation) and it supports the arguments that the duality (equivalence) is approximate in the case $N_{c}=3$.

\subsection{Predicting the Catalysis of Chiral Symmetry Breaking}\label{sec4.2}
There  have been a long debate if chiral symmetry breaking is enhanced by chiral imbalance, that is, there is catalysis of chiral symmetry breaking \cite{increase, increase11, increase111, increase1, increase2, increase3, increase4, increase5, increase55}, or it is inhibited by chiral imbalance, that is, there is anti-catalysis of chiral symmetry breaking \cite{decrease, decrease1, decrease2, decrease3, decrease4, decrease5, decrease6, decrease7, decrease8}. This question has been studied in a variety of approaches \cite{increase, increase11, increase111, increase1, increase2, increase3, increase4, increase5, increase55, decrease, decrease1, decrease2, decrease3, decrease4, decrease5, decrease6, decrease7, decrease8} and different studies came to opposite conclusions. One can say that the issue is settled after lattice simulations results \cite{Braguta,  Braguta13, Braguta1, Braguta:2016aov}, where it has been found that chiral imbalance catalyses chiral symmetry breaking in QCD. But they used unphysically large pion mass so it is still interesting to get additional indications that it is the case.

Let us discuss the possibility of using our considerations and dualities for establishing the catalysis of chiral symmetry breaking by chiral chemical potential. We have the main duality (\ref{16}) connecting isospin chemical potential $\mu_{I}$ and chiral isospin chemical potential $\mu_{I5}$. But now we talk about the influence of chiral imbalance $\mu_{5}$ on QCD phase structure, not a chiral isospin one. Nevertheless, one can note that we have another duality (\ref{19}). Using this duality one can argue that the effects of chiral isospin chemical potential $\mu_{I5}$ and chiral chemical potential $\mu_{5}$ on the phenomenon of chiral symmetry breaking are exactly the same. It~can be objected that this duality holds only if there is no pion condensation phenomenon in the system and can be broken by pion condensate. However, one~can show dynamically (at least in the framework of NJL model) that pions do not condense in the system with just chiral or chiral isospin imbalance and the condition of zero pion condensate holds in this case.
So we can use the dualities (\ref{19}), (\ref{16}) and argue that $(\mu_{5}, T)$ and $(\mu_{I}/2, T)$ phase diagrams are dual to each other if one performs the transformation PC $\leftrightarrow$ CSB. %please define the abbreviations if appropriate
  Thus one can use the duality to get the critical temperature of chiral symmetry breaking phase as a function of chiral chemical potential $T_{c}$($\mu_{5}$) from the phase structure at $\mu_{I}$. And the QCD phase structure at non-zero isospin imbalance is comparatively well-known \cite{Son, Loewe, KogutSinclair, BrandtEndrodi, BrandtEndrodi1, Adhikari:2019mlf} and one knows that the critical temperature of PC phase is an increasing function of isospin chemical potential at least to the values of several hundred MeV. One~also knows that the duality is valid only in the chiral limit ($m_{\pi}=0$) and is a very good approximation but not exact in the physical point ($m_{\pi}\approx140$ MeV) \cite{Khunjua:2018jmn} (the duality is almost exact if $\nu>m_{\pi}/2$, at~least from one hundred %100
to several hundred MeV). So one can conclude using the duality and lattice QCD results at isospin imbalance that critical temperature of chiral symmetry breaking phase should be an increasing function of chiral chemical potential $\mu_{5}$. Meaning that the catalysis of chiral symmetry breaking %by chiral chemical potential 
takes place. One can also show that the chiral condensate increases with the chiral imbalance as well and it gets harder to melt it so the critical temperature increases. This is another example of the practical use of the duality and the physical results that can be obtained from it.

\subsection{Generating the Phase Diagram without Any Calculations}\label{sec4.3}
Duality is an interesting property of the QCD phase structure in itself. But as has been shown in the previous sections one can also use it to get new results or even try to circumvent the sign problem. Here in this section we will add another example and show that it is possible to get the whole new phase diagram from duality only. 

There are a number of strong arguments supported by model calculations that at large and intermediate densities in the QCD phase diagram there exist phases with spatially inhomogeneous condensates (order parameters) (see the
reviews in References \cite{buballa,Heinz:2014pkl}).

The possible inhomogeneous phases in QCD phase structure at isospin imbalance have been studied in a number of papers.
First, it~was assumed that the pion condensate is homogeneous and only inhomogeneous chiral symmetry breaking phases are possible \cite{Nickel:2009wj,Nowakowski:2015ksa,Nowakowski,Andersen:2018osr} and the QCD-phase diagram was obtained within NJL models, for example, 
in \cite{Nickel:2009wj,Nowakowski:2015ksa,Nowakowski} (similar results was obtained in
\cite{Andersen:2018osr} in the quark- meson~model). 

It has been found that, at rather small values of isospin chemical potential $\nu=\mu_{I}/2$ ($\nu< 60$ MeV) and for $\mu\gtrsim 300$ MeV, there might appear a region of 
inhomogeneous chiral symmetry breaking (ICSB) phase. The~considerations were performed in the chiral limit for simplicity. But one can think that it is a good approximation because the case with zero isospin %asymmetry
imbalance in physical point was considered 
in Reference~\cite{Nickel:2009wj} and it was shown that  qualitative picture does not change with the non-zero current quark mass. With increase of the mass %the critical point at the 
%$(\mu,T)$-phase diagram just shifts to smaller values of temperature and larger chemical potentials and 
the region of the inhomogeneous phase only gets smaller in size. Probably the influence of quark mass stays qualitatively the same also in the case of non-zero isospin chemical~potential.

Second, it~was assumed in Reference \cite{he} that chiral condensate is homogeneous and a spatially inhomogeneous charged pion condensation (ICPC) phase has been studied in the framework of NJL model. It~was noted that inhomogeneous pion condensation phase is realized in the phase diagram at rather high isospin chemical potential $\nu\gtrsim 400$ MeV.  

\begin{figure}
\centering
\includegraphics[width=0.65\textwidth]{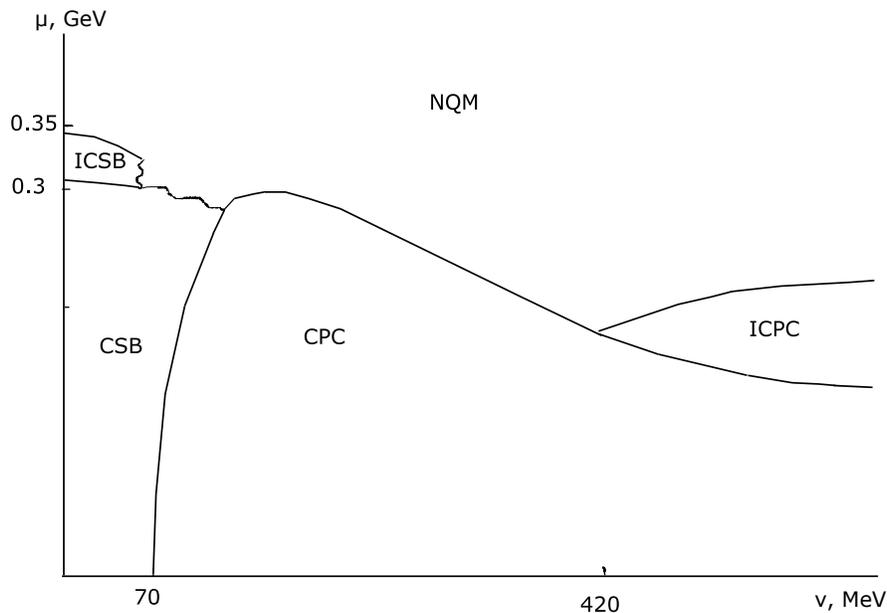}
\caption{The ($(\nu, \mu)$-phase diagram at $\mu_{5}=\nu_5=0$.}
\label{fig1}
\end{figure}

It is possible to connect these situations and obtain the full $(\nu, \mu)$ phase diagram assuming the possibility of both inhomogeneous charged pion condensation and chiral symmetry breaking phases. It~is possible due to the fact that the regions of inhomogeneous phases of different studies do not overlap at the phase diagram and the assumption that there is no mixed inhomogeneous phase. The~latter assumption in principle can be lifted and what we can get is only a more rich picture of inhomogeneous phases.  
So one can envisage the full schematic $(\nu, \mu)$-phase portrait of quark matter with baryon density and isospin 
imbalance, it~is shown in Figure~\ref{fig1}. 

\begin{figure}
\centering
\includegraphics[width=0.65\textwidth]{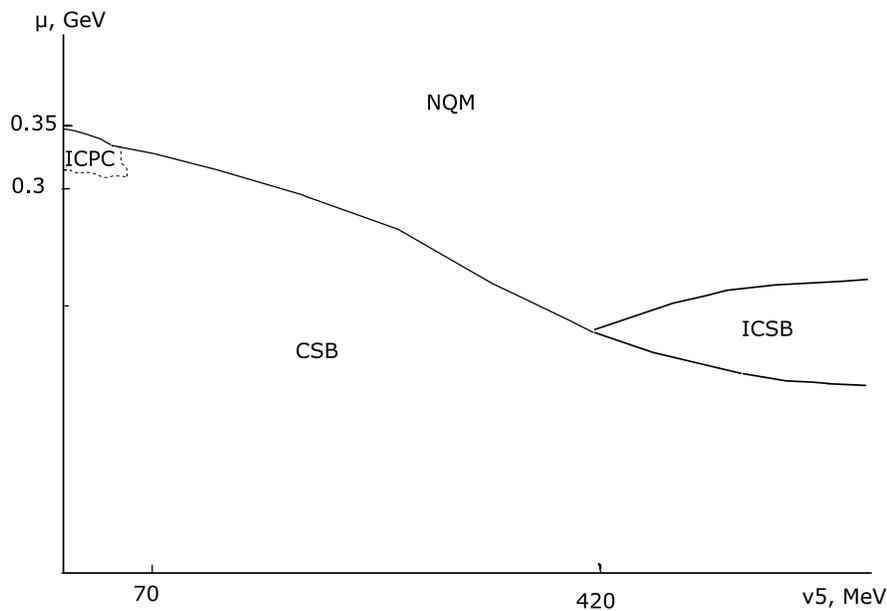}
\caption{The ($(\nu_{5}, \mu)$-phase diagram at $\mu_{5}=\nu=0$.}
\label{fig2}
\end{figure}

Now one can note that the phase diagram $(\nu, \mu)$ that we have discussed above can be transformed by the main duality (\ref{16}) to the phase diagram $(\nu_{5}, \mu)$. And~we get two  phase diagrams $(\nu, \mu)$ and $(\nu_{5}, \mu)$ that are dual to each other. As was discussed above, the first one is more or less known at least in effective models but the latter is completely unexplored and one has no idea how it looks like.
And by this dual transformation one can get %the idea about
completely yet not considered part of QCD phase diagram, namely, $(\nu_{5}, \mu)$ phase diagram that is depicted in Figure~\ref{fig2}. One can see that there is ICPC phase in the region corresponding to rather high values of $\mu$ and, probably, baryon density is non-zero in this region.
It can also be noticed that there is ICSB phase at values of chemical potential $\mu$ around 200 MeV and rather large chiral isospin chemical potential $\nu_{5}$ (see Figure~\ref{fig2}). Due to rather large chiral isospin chemical potential a part of this region might have non-zero baryon density. One can conclude that in inhomogeneous case phase diagram seems to be rather rich and (dense) quark matter with chiral isospin imbalance can have various inhomogeneous condensates, namely chiral or charged pion one.

The above has used the duality in inhomogeneous case but it is far from obvious that the duality is valid in this case as well. However, it~has been demonstrated in Reference \cite{Khunjua:2019lbv} that it is the case. It~is rather nontrivial fact and it indicates that probably the duality is a more deep property of the QCD phase diagram.
%It was shown that dense quark matter with isospin imbalance is in a sense dual to dense (non-zero baryon density) quark matter with chiral imbalance. %Dense means non-zero baryon density 
Let us also stress that the duality holds for the case of non-zero baryon density and it is, in~particular, an interesting feature of dense quark matter.
%and this fact follows from the fact that duality holds at non-zero baryon density as well.

%%%%%%%%%%%%%%%%%%%%%%%%%%%%%%%%%%%%%%

Let us also note that at high baryon density a different phenomenon are expected to %sets in
take place. At~low temperatures and sufficiently large baryon chemical potential the colour interaction (in the color anti-symmetric channel) starts to favour the formation of non-zero %quark quark
diquark (quark-quark) %condensate) 
condensate~\cite{supercond}. Due to the fact that this phenomenon breaks colour symmetry it is called colour superconductivity. There could be also interesting inhomogeneous structure of condensates \cite{inhomsupercond, Anglani:2013gfu}. Throughout this paper we neglect the possibility of colour superconductivity phase but it is also very interesting to study the dualities if it takes place. However, it~is outside the scope of this paper and we leave it for the future.
%\cite{supercond} \cite{inhomsupercond}

%%%%%%%%%%%%%%%%%%%%%%%%%%%%%%%%%%%%%%%%5

\section{Conclusions}\label{sec5}

The dualities of the QCD phase diagram, in particular, the~duality between chiral symmetry breaking and pion condensation phenomena has been found in the framework of the (1+1)-dimensional QCD motivated toy model in Reference \cite{kkzz, %kkzzp,
Khunjua:2019ini, Ebert:2016hkd, Khunjua:2018vyb}. Then it was shown to exist in the framework of effective models in References~\cite{kkz3, kkz4, Khunjua:2018jmn, Khunjua:2019lbv, Khunjua:2018ant, Khunjua:2018xly, Khunjua:2019ojm}.

In this paper we have endeavoured to show that the duality is not just an interesting mathematical fact in itself and an interesting feature intrinsic to the phase diagram of dense quark matter (that it surely is) but also a powerful tool that can be used to produce new results %with immense
almost effortlessly. There~are even ideas that it can help (not solve but circumvent) the sign problem (see Section~\ref{sec4.1}).
Moreover, it~is known that there is a contradiction between the predictions of different studies of the influence of chiral imbalance on chiral symmetry breaking phenomenon. Some works predicted that there should be catalysis of chiral symmetry breaking, others that there is anti-catalysis.   It is shown in the framework of duality that it is possible to, if not settle the issue completely, %with contradiction of %with catalysis of chiral symmetry breaking
 surely make a strong argument to favour the existence of catalysis of chiral symmetry breaking. %phenomenon.

Another argument, an even more trustworthy one, is the lattice simulations \cite{Braguta, Braguta13, Braguta1, Braguta:2016aov} (first principle approach) that, however, performed not at physical pion mass, gives a decisive answer to this question. %}
One can probably argue that our results, combined with the lattice simulations, %(not at physical pion mass) 
can claim that there is not a lot of doubt that this effect indeed takes place. Then we showed that the duality can be used as a tool for plotting entirely new phase diagrams completely for free in terms of efforts. It~is demonstrated by constructing the %($\mu, \nu_{5}$)
phase diagram of dense quark matter with chiral imbalance.

The basic features of the GN model and its extensions, including the question of why it might be interesting in the context of QCD, are summarised at the beginning of the paper.
Then it is shown how to obtain the duality property with different approaches (including the above-mentioned toy model). After~that, the picture with several additional dualities of dense quark matter has been discussed. Eventually, the~possible applications of dualities have been considered.

Let us enlist the main %employments 
applications
of dualities that have been discussed in this paper.

\begin{itemize}
\item There has been discussed the possibility of circumventing the sign problem by constructing dualities between QCD phase diagrams with different chemical potentials. %s been expanded to our dualities and it is argued that it is a feasible scenario.

\item  It is shown that a proble,m if there exists catalysis or anti-catalysis of chiral symmetry breaking by chiral imbalance, can be resolved just by duality property to the favour of catalysis. And bearing also in mind the lattice simulations results at unphysically large pion mass one can say that there is not much doubt that this issue is~settled.    %the rather well-established knowledge of pion condensation properties at isospin density.

\item	The whole new phase diagram of dense quark matter with chiral imbalance with the possibility of different inhomogeneous phases has been obtained just by duality only and previously known~results. %It is shown that the duality can be used to produce new results, new phase diagrams (different sections of phase diagram). As an example, it~is showcased how from the phase structure of dense quark matter with non-zero isospin density (including possibility of inhomogeneous  condensates (phases)) one can obtain, based on the duality only, the~phase structure of dense quark matter with chiral imbalance.
\end{itemize}

So the dualities can be used and can be very helpful in understanding the phase structure of QCD, including the large baryon density region.

%%%%%%%%%%%%%%%%%%%%%%%%%%%%%%%%%%%%%%%%%%%%

Let us make another note on the possible applications of dualities in astrophysics.
 Dense matter with isospin imbalance can be easily found inside neutron stars. Chirality can be probably generated in heavy ion collisions, for example, due to strong electromagnetic fields (see introduction). It~is demonstrated in this paper that (dense) matter with isospin imbalance is connected by duality with (dense) matter with chiral imbalance (chiral isospin chemical potential). So maybe one can think that using the main duality, phenomena in cores of neutron stars can be probed in the terrestrial heavy ion collision experiments. Besides, since in neutron stars the baryon density is rather high (huge) and main duality leave baryon density intact, one needs at the other side large baryon density in heavy ion collisions, which is possible only at not so high energy, for~example, as at NICA complex or other projects discussed in the introduction. It~is a rather interesting opportunity but there are a number of hindrances. For~example, the~conditions in this settings are different as in neutron stars there should be, for~example, $\beta$-equilibrium condition. Also, since duality does not change temperature (let us note that it has been shown in Reference \cite{Khunjua:2018jmn} that the duality is valid also in the case of non-zero or even high temperatures), at both sides of possibly connected by duality phenomena %cases
 there should be similar temperatures. And even in the intermediate energy heavy ion collision experiments one talks about rather large temperatures that is not even closely realized in individual neutron stars (even in proto-neutron stars, where temperatures can reach 10 MeV, they are still %the temperatures is
 smaller). 
 But here one can think about %supervona explosions and 
 recently observed mergers of neutron stars \cite{TheLIGOScientific:2017qsa}.
 Since in neutron star mergers the temperature can reach %quite high 
 values %, e.g., 
 as high as %it is  approximately 
80 MeV %can be reached 
\cite{Galeazzi:2013mia, Dexheimer:2017nse} %(Galeazzi et al. (2013)).
%These temperatures 
and they are not significantly different
from the ones reached in intermediate energy heavy-ion collisions ($\beta$-equilibrium condition in this case is also slightly different from cold neutron star case~\cite{Alford:2018lhf}), %, especially moderate energy ones. %the neutron star mergers
it is a more plausible candidate to be mapped  by duality to heavy ion collisions.
Also let us note that supernova explosions, where temperatures can be rather high \cite{Fischer:2010wp}, and matter during the black hole formation from a gravitational collapse of a massive star, where temperatures could be even higher ($T\sim 90$ MeV \cite{Ohnishi:2011jv} or even over 100~MeV~%higher
\cite{Sumiyoshi:2006id}), are %is 
also viable for this role. %candidate
%But still maybe this two completely different settings can be somehow mapped into each other and interesting inference can be made then. 
If one assumes that all the above conditions are fulfilled, then the conditions dual to %phase structure of matter
the ones during neutron star mergers can be realized and studied at intermediate energy heavy ion collision experiments such as NICA.
It is also even more feasible to get interesting information by duality connecting baryon chemical potential (with the another one) that we talked about in Section~\ref{sec4.1}, especially try to get information about equation of state from phase structure of QCD at non-zero isospin or chiral imbalances. It~can be studied in the future.
\vspace{6pt} 

%%%%%%%%%%%%%%%%%%%%%%%%%%%%%%%%%%%%%%%%%%
%% optional
%\supplementary{The following are available online at \linksupplementary{s1}, Figure S1: title, Table S1: title, Video S1: title.}


%%%%%%%%%%%%%%%%%%%%%%%%%%%%%%%%%%%%%%%%%%

%%%%%%%%%%%%%%%%%%%%%%%%%%%%%%%%%%%%%%%%%%

%%%%%%%%%%%%%%%%%%%%%%%%%%%%%%%%%%%%%%%%%%
\section{Acknowledgments} R.N.Z. is grateful for support of Russian Science Foundation under the grant No 19-72-00077. The work is also supported by the Foundation for the Advancement of Theoretical Physics and Mathematics
BASIS.

%The authors would like to thank the organizers of ``The II International Workshop on Theory of Hadronic Matter Under Extreme Conditions'' {Victor V. Braguta, Evgeni E. Kolomeitsev, David Blaschke, Sergei N. Nedelko, Alexandra V. Friesen, Vladimir E. Voronin, Olga N. Belova }for a very fruitful workshop.}%please provide the full names of the persons.

%%%%%%%%%%%%%%%%%%%%%%%%%%%%%%%%%%%%%%%%%%

%%%%%%%%%%%%%%%%%%%%%%%%%%%%%%%%%%%%%%%%%%
%% optional
%\abbreviations{The following abbreviations are used in this manuscript:\\

%\noindent 
%\begin{tabular}{@{}ll}
%TDP & thermodynamic potential\\
%GN model & Gross-Neveu model\\
%NJL model & Nambu--Jona-Lasinio model\\
%CSB & chiral symmetry breaking\\
%PC & pion condensation\\
%CPC & charged pion condensation\\
%ICSB & inhomogeneous chiral symmetry breaking\\
%ICPC & inhomogeneous charged pion condensation\\
%\end{tabular}}

%%%%%%%%%%%%%%%%%%%%%%%%%%%%%%%%%%%%%%%%%%
%% optional
%\appendixtitles{no} %Leave argument "no" if all appendix headings stay EMPTY (then no dot is printed after "Appendix A"). If the appendix sections contain a heading then change the argument to "yes".


%=====================================
% References, variant A: internal bibliography
%=====================================
%\reftitle{References}

%%%%%%%%%%%%%%%%%%%%%%%%%%%%%%%%%%%%%%%%%%

\begin{thebibliography}{999}

\bibitem{Mannarelli}
Mannarelli, M.
Meson Condensation.
\emph{Particles} {\bf 2019}, \emph{2}, 411--443.

\bibitem{Ayala:2011vs}
Ayala, A.; Bashir, A.; Dominguez, C.A.; Gutierrez, E.; Loewe, M.; Raya, A.
"QCD phase diagram from finite energy sum rules."
\emph{Phys.\ Rev.\ D} {\bf 2011}, \emph{84}, 056004.%There are more than one Ref. in the same Reference, please check if the highlight part can be deleted. If not, please name it as another ref and complete necessary information  (authors' names and aritcile titles)..
  %doi:10.1103/PhysRevD.84.056004
  %[arXiv:1106.5155 [hep-ph]];
  %%CITATION = doi:10.1103/PhysRevD.84.056004;%%
  %37 citations counted in INSPIRE as of 02 Dec 2019
  
  \bibitem{Ayala1}
Ayala, A.; Bashir, A.; Cobos-Martínez, J.J.; Hernández-Ortiz, S.; Raya, A.
"The effective QCD phase diagram and the critical end point."
\emph{Nucl. Phys. B} {\bf2015}, 897.

  %\cite{Hayashi:2010ck}
\bibitem{Inagaki}
Hayashi, M.; Inagaki, T.; Sakamoto, W.
Phase Structure of a Four and Eight-Fermion Interaction Model at Finite Temperature and Chemical Potential in Arbitrary Dimensions.
\emph{Int.\ J.\ Mod.\ Phys.\ A} {\bf 2010}, \emph{25}, 4757.%There are more than one Ref. in the same Reference, please check if the highlight part can be deleted. If not, please name it as another ref and complete necessary information (authors' names and aritcile titles).
  %doi:10.1142/S0217751X10050627
  %[arXiv:1007.1497 [hep-ph]];
  %%CITATION = doi:10.1142/S0217751X10050627;%%
  %3 citations counted in INSPIRE as of 04 Dec 2019
  %\cite{Fujihara:2008ae}
\bibitem{Fujihara:2008ae}
  Fujihara, T.; Kimura, D.; Inagaki, T.; Kvinikhidze, A.
  "High density quark matter in the NJL model with dimensional vs. cut-off regularization."
  \emph{Phys.\ Rev.\ D} {\bf 2009}, \emph{79}, 096008.
  %doi:10.1103/PhysRevD.79.096008
  %[arXiv:0812.2821 [hep-ph]];
  %%CITATION = doi:10.1103/PhysRevD.79.096008;%%
  %29 citations counted in INSPIRE as of 04 Dec 2019
  
  %\cite{Fujihara:2008wx}
\bibitem{Fujihara:2008wx}
 Fujihara,  T.; Inagaki,  T.; Kimura,  D.; A.~Kvinikhidze,
  "Reconsideration of the 2-flavor NJL model with dimensional regularization at finite temperature and density."
  \emph{Prog.\ Theor.\ Phys.\ Suppl.} {\bf 2008}, \emph{174}, 72.
  %doi:10.1143/PTPS.174.72
  %[arXiv:0806.1331 [hep-ph]].
  %%CITATION = doi:10.1143/PTPS.174.72;%%
  %7 citations counted in INSPIRE as of 04 Dec 2019
  
  %\cite{Friesen:2014mha}
\bibitem{Friesen:2014mha}
 Friesen,  A.V.; Kalinovsky, Y.L.; Toneev, V.D.
 Vector interaction effect on thermodynamics and phase structure of QCD matter.
  \emph{Int.\ J.\ Mod.\ Phys.\ A}  {\bf 2015}, \emph{30},  1550089.
  %doi:10.1142/S0217751X1550089X
  %[arXiv:1412.6872 [hep-ph]].
  %%CITATION = doi:10.1142/S0217751X1550089X;%%
  %15 citations counted in INSPIRE as of 28 Nov 2019
  
  %\cite{Nedelko:2014sla}
\bibitem{Nedelko:2014sla}
  Nedelko, S.N.; Voronin,  V.E.
  Domain wall network as QCD vacuum and the chromomagnetic trap formation under extreme conditions.
  \emph{Eur.\ Phys.\ J.\ A}  {\bf 2015}, \emph{51},  45.
 % doi:10.1140/epja/i2015-15045-8
  %[arXiv:1403.0415 [hep-ph]].
  %%CITATION = doi:10.1140/epja/i2015-15045-8;%%
  %10 citations counted in INSPIRE as of 28 Nov 2019
  
  %\cite{Blaschke:2019tbh}
\bibitem{Blaschke}
  Blaschke, D.; Alvarez-Castillo, D.E.; Ayriyan, A.; Grigorian, H.; Lagarni, N.K.; Weber, F.
  "Astrophysical aspects of general relativistic mass twin stars."
  \emph{arXiv} {\bf2019}, arXiv:1906.02522.%There are more than one Ref. in the same Reference, please check if the highlight part can be deleted.
  %%CITATION = ARXIV:1906.02522;%%
  
  \bibitem{Shahrbaf:2019vtf}
  Shahrbaf, M.; Blaschke, D.; Grunfeld,  A.G.; Moshfegh, H.R.
  "First-order phase transition from hypernuclear matter to deconfined quark matter obeying new constraints from compact star observations."
  \emph{arXiv} {\bf2019}, arXiv:1908.04740.
  %%CITATION = ARXIV:1908.04740;%%
  
 \bibitem{Bauswein:2019skm}
  Bauswein, A.; et al.
  "Equation-of-state Constraints and the QCD Phase Transition in the Era of Gravitational-Wave Astronomy."
  AIP Conf. Proc.  {\bf 2019}, \emph{2127},  020013.
  %doi:10.1063/1.5117803
  %[arXiv:1904.01306 [astro-ph.HE]];
  %%CITATION = doi:10.1063/1.5117803;%%
  %8 citations counted in INSPIRE as of 28 Nov 2019
  
  \bibitem{Alvarez-Castillo:2018rrv}
  Alvarez-Castillo, D.; Blaschke, D.
  "A Mixing Interpolation Method to Mimic Pasta Phases in Compact Star Matter."
  \emph{arXiv} {\bf2018}, arXiv:1807.03258.
  %%CITATION = ARXIV:1807.03258;%%
  %1 citations counted in INSPIRE as of 28 Nov 2019
  
  \bibitem{Radzhabov:2010dd}
 Radzhabov, A.E.;  Blaschke, D.; Buballa, M.;  Volkov, M.K.
  "Nonlocal PNJL model beyond mean field and the QCD phase transition."
  \emph{Phys.\ Rev.\ D} {\bf 2011}, \emph{83}, 116004.
  %doi:10.1103/PhysRevD.83.116004
 % [arXiv:1012.0664 [hep-ph]].
  %%CITATION = doi:10.1103/PhysRevD.83.116004;%%
  %65 citations counted in INSPIRE as of 28 Nov 2019
  
  %\cite{Rajagopal:1999cp}
\bibitem{Rajagopal:1999cp}
  Rajagopal, K.
  Mapping the QCD phase diagram.
  \emph{Nucl.\ Phys.\ A}  {\bf 1999}, \emph{661}, 150.
  %doi:10.1016/S0375-9474(99)85017-9
 % [hep-ph/9908360].
  %%CITATION = doi:10.1016/S0375-9474(99)85017-9;%%
  %101 citations counted in INSPIRE as of 28 Nov 2019
  
  %\cite{Tawfik:2019tkp}
\bibitem{Tawfik:2019tkp}
  Tawfik, A.N.; Diab, A.M.; Ghoneim, M.T.; Anwer, H.
  SU(3) Polyakov Linear-Sigma Model With Finite Isospin Asymmetry: QCD Phase Diagram.
  \emph{Int.\ J.\ Mod.\ Phys.\ A}  {\bf 2019}, \emph{34}, 1950199.
  %doi:10.1142/S0217751X19501999
  %[arXiv:1904.09890 [hep-ph]].
  %%CITATION = doi:10.1142/S0217751X19501999;%%
  %3 citations counted in INSPIRE as of 02 Dec 2019
 
  %\cite{Sasaki:2009zh}
\bibitem{Sasaki:2009zh}
  Sasaki, C.
  The QCD Phase Diagram from Chiral Approaches.
  \emph{Nucl.\ Phys.\ A} {\bf 2009},  \emph{830},  649.
  
  %\cite{Grigorian:2017xqd}
\bibitem{Grigorian:2017xqd}
  Grigorian, H.; Kolomeitsev, E.E.; Maslov, K.A.; Voskresensky, D.N.
  On Cooling of Neutron Stars with a Stiff Equation of State Including Hyperons.
  \emph{Universe} {\bf 2018},  \emph{4},  29.%There are more than one Ref. in the same Reference, please check if the highlight part can be deleted.
 % doi:10.3390/universe4020029
 % [arXiv:1801.00040 [astro-ph.HE]].
  %%CITATION = doi:10.3390/universe4020029;%%
  %2 citations counted in INSPIRE as of 06 Dec 2019
  %\cite{Kolomeitsev:2017gli}
\bibitem{Kolomeitsev:2017gli}
  E.~E.~Kolomeitsev, K.~A.~Maslov and D.~N.~Voskresensky,
  "Charged $\rho$-meson condensation in neutron stars."
  \emph{Nucl.\ Phys.\ A}  {\bf 2018}, \emph{970}, 291.
 % doi:10.1016/j.nuclphysa.2017.12.002
%  [arXiv:1710.06749 [nucl-th]].
  %%CITATION = doi:10.1016/j.nuclphysa.2017.12.002;%%
  %6 citations counted in INSPIRE as of 06 Dec 2019
  
\bibitem{NICAWhitePaper}
Blaschke, D., Aichelin, J., Bratkovskaya, E.; Friese, V.; Gazdzicki, M.; Randrup, J.; Rogachevsky, O.; Teryaev, O.;  Toneev, V.
"Topical issue on exploring strongly interacting matter at high densities-nica white paper."
\emph{Eur.~Phys. J. A}  {\bf 2016}, \emph{52}, 267. %https://doi.org/10.1140/epja/i2016-16267-x

\bibitem{KogutSinclair} 
Kogut, J.B.; Sinclair, D.K.
  "Quenched lattice QCD at finite isospin density and related theories."
  \emph{ Phys.\ Rev.\ D} {\bf 2002}, \emph{66}, 014508.\\%There are more than one Ref. in the same Reference, please check if the highlight part can be deleted.
  %doi:10.1103/PhysRevD.66.014508
  %[hep-lat/0201017].
  %%CITATION = doi:10.1103/PhysRevD.66.014508;%% 
 "Lattice QCD at finite isospin density at zero and finite temperature."
  \emph{ Phys.\ Rev.\ D} {\bf 2002}, \emph{66}, 034505.\\
  %doi:10.1103/PhysRevD.66.034505
  %[hep-lat/0202028].
  %%CITATION = doi:10.1103/PhysRevD.66.034505;%% 
"The Finite temperature transition for 2-flavor lattice QCD at finite isospin density."
  \emph{ Phys.\ Rev.\ D} {\bf 2004}, \emph{70}, 094501.
  %doi:10.1103/PhysRevD.70.094501
  %[hep-lat/0407027].
  %%CITATION = doi:10.1103/PhysRevD.70.094501;%%
  
  \bibitem{BrandtEndrodi} 
 Brandt, B.B.; Endrodi, G.; Schmalzbauer, S.
  "QCD phase diagram for nonzero isospin-asymmetry."
  \emph{ Phys.\ Rev.\ D} {\bf 2018}, \emph{97}, 054514.\\
  "QCD at finite isospin chemical potential."
  \emph{EPJ Web Conf.}   {\bf 2018}, \emph{175}, 07020.
  %doi:10.1103/PhysRevD.97.054514
  %[arXiv:1712.08190 [hep-lat]].
  %%CITATION = doi:10.1103/PhysRevD.97.054514;%%
  
  \bibitem{BrandtEndrodi1} 
  Brandt, B.B.; Endrodi, G.
  "QCD phase diagram with isospin chemical potential."
  \emph{PoS LATTICE} {\bf 2016}, \emph{2016}, 039.
  
  %\cite{Schwarz:2009ii}
\bibitem{Schwarz:2009ii}
 Schwarz,  D.~J.; Stuke, M.
  "Lepton asymmetry and the cosmic QCD transition."
  \emph{J. Cosmol. Astropart. Phys.}  {\bf 2009}, \emph{ 0911}, 025;
   Erratum: \emph{J. Cosmol. Astropart. Phys.}  {\bf 2010}, \emph{1010}, E01.
   %The journal's information is complete, please check if it can be deleted.
  %doi:10.1088/1475-7516/2009/11/025, 10.1088/1475-7516/2010/10/E01
  %[arXiv:0906.3434 [hep-ph]].
  %%CITATION = doi:10.1088/1475-7516/2009/11/025, 10.1088/1475-7516/2010/10/E01;%%
  %76 citations counted in INSPIRE as of 28 Nov 2019
  
    \bibitem{Metlitski} 
  Metlitski, M.~A.; Zhitnitsky, A.~R.
  Anomalous axion interactions and topological currents in dense matter.
  \emph{ Phys.\ Rev.\ D} {\bf 2005}, \emph{72}, 045011.
%  doi:10.1103/PhysRevD.72.045011
 % [hep-ph/0505072].
  %%CITATION = doi:10.1103/PhysRevD.72.045011;%%
  
\bibitem{Fukushima:2018grm} 
  Fukushima, K.
  "Extreme matter in electromagnetic fields and rotation."
  \emph{Prog.\ Part.\ Nucl.\ Phys.} {\bf 2019}, \emph{107}, 167.
  %%CITATION = doi:10.1016/j.ppnp.2019.04.001;%%
  %12 citations counted in INSPIRE as of 06 Dec 2019
 % arXiv:1812.08886 [hep-ph].
  %%CITATION = ARXIV:1812.08886;%%  
  
   \bibitem{Khunjua:2018jmn}
  Khunjua, T.G.; Klimenko, K.G.; Zhokhov, R.N.
  "Chiral imbalanced hot and dense quark matter: NJL analysis at the physical point and comparison with lattice QCD."
 \emph{Eur.\ Phys.\ J.\ C}  {\bf 2019}, \emph{79}, 151.%There are more than one Ref. in the same Reference, please check if the highlight part can be deleted.
  %doi:10.1140/epjc/s10052-019-6654-2
  %[arXiv:1812.00772 [hep-ph]].
  %%CITATION = doi:10.1140/epjc/s10052-019-6654-2;%% 
  
  %\cite{Khunjua:2018dbm}
\bibitem{Khunjua:2018dbm}
  T.~G.~Khunjua, K.~G.~Klimenko and R.~N.~Zhokhov,
  ``QCD phase diagram with chiral imbalance in NJL model: duality and lattice QCD results.
  \emph{J. Phys. Conf. Ser.} {\bf 2019}, \emph{1390}, 012015.
 % arXiv:1812.01392 [hep-ph].
  %%CITATION = ARXIV:1812.01392;%%
  \bibitem{KhunjuaMoscowUnivBull}
  Khunjua, T.~G.; Klimenko, K.~G. and Zhokhov, R.~N.
  "Pion Condensation in Hot Dense Quark Matter with Isospin and Chiral-Isospin Asymmetries within the Nambu—Jona-Lasinio Model."
  \emph{Moscow Univ.\ Phys.\ Bull.} {\bf2019}, \emph{ 74}, no.5,  473.
   %\cite{Khunjua:2019ojm}
%\bibitem{Khunjua:2019ojm}
%  T.~G.~Khunjua, K.~G.~Klimenko and R.~N.~Zhokhov,
  %``Quark/hadronic matter and dualities of QCD thermodynamics.
  %arXiv:1912.09102 [hep-ph].
  %%CITATION = ARXIV:1912.09102;%%

  \bibitem{Ruggieri:2016fny}
Ruggieri, M.; Peng, G.~X.; Chernodub, M.
"Chiral medium produced by parallel electric and magnetic fields."
\emph{EPJ Web Conf.} {\bf 2016}, \emph{129}, 00037.\\
"Chiral relaxation time at the crossover of quantum chromodynamics."%There are more than one Ref. in the same Reference, please check if the highlight part can be deleted.
\emph{Phys.\ Rev.\ D} {\bf 2016}, \emph{94},  054011.
%doi:10.1051/epjconf/201612900037
%[arXiv:1609.04537 [hep-ph]];
%%CITATION = doi:10.1051/epjconf/201612900037;%%

  \bibitem{RuggieriPeng}
Ruggieri, M.; Lu, Z.Y.; Peng, G.X.
"Influence of chiral chemical potential, parallel electric, and magnetic fields on the critical temperature of QCD."
\emph{Phys.\ Rev.\ D} {\bf 2016}, \emph{94},  116003.
%doi:10.1103/PhysRevD.94.116003
%[arXiv:1608.08310 [hep-ph]];
%%CITATION = doi:10.1103/PhysRevD.94.116003;%%
%M.~Ruggieri, G.~X.~Peng and M.~Chernodub,
%``Chiral Relaxation Time at the Crossover of Quantum Chromodynamics.
%\emph{ Phys.\ Rev.\ D} {\bf 94},  054011 (2016); \emph{ Phys.\ Rev.\ D} {\bf 93}, 094021 (2016).
%doi:10.1103/PhysRevD.94.054011
%[arXiv:1606.03287 [hep-ph]];
%%CITATION = doi:10.1103/PhysRevD.94.054011;%%

\bibitem{Ruggieri:2016fny1}
Ruggieri, M.; Peng, G.X.
"Quark matter in a parallel electric and magnetic field background: Chiral phase transition and equilibration of chiral density."
\emph{Phys.\ Rev.\ D} {\bf 2016}, \emph{93}, 094021.
%doi:10.1103/PhysRevD.93.094021
%[arXiv:1602.08994 [hep-ph]].
%%CITATION = doi:10.1103/PhysRevD.93.094021;%%

  \bibitem{Braguta} 
 Braguta,  V.~V.; Goy, V.~A.; Ilgenfritz, E.-M.; Kotov, A.~Y.; Molochkov, A.~V.; Muller-Preussker, M.; Petersson, B.
  "Two-Color QCD with Non-zero Chiral Chemical Potential."
  \emph{ J. High Energy Phys.} {\bf 2015}, \emph{1506}, 094.%There are more than one Ref. in the same Reference, please check if the highlight part can be deleted.
 % doi:10.1007/JHEP06(2015)094
  %[arXiv:1503.06670 [hep-lat]].
  %%CITATION = doi:10.1007/JHEP06(2015)094;%%
  
   \bibitem{Braguta13}
Braguta,  V.V.; Ilgenfritz, E.M.; Kotov, A.Y.; Petersson, B.; Skinderev, S.A.
  "Study of QCD Phase Diagram with Non-Zero Chiral Chemical Potential."
  \emph{ Phys.\ Rev.\ D} {\bf 2016}, \emph{93}, 034509.
%  doi:10.1103/PhysRevD.93.034509
 % [arXiv:1512.05873 [hep-lat]].
  %%CITATION = doi:10.1103/PhysRevD.93.034509;%%
  
   \bibitem{Braguta1}
Braguta, V.V.; Ilgenfritz, E.M.; Kotov, A.Y.; Muller-Preussker, M.; Petersson, B.; Schreiber, A.
  "Two-Color QCD with Chiral Chemical Potential."
  \emph{PoS LATTICE} {\bf 2014}, \emph{235}.
%  doi:10.22323/1.214.0235
 % [arXiv:1411.5174 [hep-lat]].
  %%CITATION = doi:10.22323/1.214.0235;%%
  
 \bibitem{Braguta:2016aov}
 Braguta, V.V.; Kotov, A.Y.
  "Catalysis of Dynamical Chiral Symmetry Breaking by Chiral Chemical Potential."
  \emph{Phys.\ Rev.\ D}   {\bf 2016}, \emph{ 93}, 105025.
 % doi:10.1103/PhysRevD.93.105025
 % [arXiv:1601.04957 [hep-th]].
  %%CITATION = doi:10.1103/PhysRevD.93.105025;%%
  %30 citations counted in INSPIRE as of 04 Dec 2018
  
  \bibitem{andrianov}
 Andrianov, A.A.; Espriu,  D.; Planells, X.
  An effective QCD Lagrangian in the presence of an axial chemical potential.
  \emph{Eur.\ Phys.\ J.\ C}  {\bf 2013}, \emph{73}, 2294.%There are more than one Ref. in the same Reference, please check if the highlight part can be deleted.
  %%CITATION = ARXIV:1210.7712;%%
  
   \bibitem{GattoRuggieri}
Gatto, R.; Ruggieri, M.
  "Hot Quark Matter with an Axial Chemical Potential."
  \emph{Phys.\ Rev.\ D} {\bf 2012}, \emph{85}, 054013.
  %%CITATION = ARXIV:1110.4904;%%
  
    \bibitem{Liu}
 Yu, L.; Liu, H.; Huang, M.
  "Spontaneous generation of local CP violation and inverse magnetic catalysis."
 \emph{Phys.\ Rev.\ D} {\bf 2014}, \emph{90}, 074009.
  %%CITATION = doi:10.1103/PhysRevD.90.074009;%%
  
    \bibitem{Liu4}
Yu, L.; Liu, H.; Huang, M.
"Effect of the chiral chemical potential on the chiral phase transition in the NJL model with different regularization schemes."
 \emph{ Phys.\ Rev.\ D} {\bf 2016}, \emph{94}, 014026.
  %%CITATION = ARXIV:1511.03073;%%
%G.~Cao and P.~Zhuang,
  %``Effects of chiral imbalance and magnetic field on pion superfluidity and color superconductivity.
 % \emph{ Phys.\ Rev.\ D} {\bf 92}, 105030 (2015);
 % %%CITATION = doi:10.1103/PhysRevD.92.105030;%%
 
  \bibitem{RuggieriandPeng}
Ruggieri M.; Peng, G.X.
  "Critical Temperature of Chiral Symmetry Restoration for Quark Matter with a Chiral Chemical Potential."
  \emph{J.\ Phys.\ G}  {\bf 2016}, \emph{43}, 125101.
  %doi:10.1088/0954-3899/43/12/125101
  %[arXiv:1602.05250 [hep-ph]].
  %%CITATION = doi:10.1088/0954-3899/43/12/125101;%%
  
    \bibitem{Zhuang}
Cao, G.; Zhuang, P.
"Effects of chiral imbalance and magnetic field on pion superfluidity and color superconductivity."
\emph{ Phys.\ Rev.\ D} {\bf 2015} \emph{92}, 105030.
%doi:10.1103/PhysRevD.92.105030
%[arXiv:1505.05307 [nucl-th]].
%%CITATION = doi:10.1103/PhysRevD.92.105030;%%
%\cite{Suenaga:2019jqu}

\bibitem{Suenaga:2019jqu}
  Suenaga, D.; Suzuki, K.; Araki, Y.; Yasui, S.
  "Kondo effect driven by chirality imbalance."
  \emph{arXiv} {\bf 2019}, arXiv:1912.12669.
  %%CITATION = ARXIV:1912.12669;%%
  
    %\cite{Gasser:1986vb}
    \bibitem{GasserLeutwyler}
    Gasser, J.; Leutwyler, H.
    Light Quarks at Low Temperatures.
    \emph{Phys.\ Lett.\ B} {\bf 1987}, \emph{184}, 83.\\%There are more than one Ref. in the same Reference, please check if the highlight part can be deleted.
    %  doi:10.1016/0370-2693(87)90492-8
    %%CITATION = doi:10.1016/0370-2693(87)90492-8;%%
    %665 citations counted in INSPIRE as of 24 Dec 2019
    %\cite{Gasser:1987ah}
    %\bibitem{Gasser:1987ah}
    %  J.~Gasser and H.~Leutwyler,
    ``Thermodynamics of Chiral Symmetry."
    \emph{Phys.\ Lett.\ B} {\bf1987}, \emph{ 188}, 477.
    %  doi:10.1016/0370-2693(87)91652-2
    %%CITATION = doi:10.1016/0370-2693(87)91652-2;%%
    %486 citations counted in INSPIRE as of 24 Dec 2019
    
      %\cite{Florkowski:1996wf}
      \bibitem{Florkowski:1996wf}
      Florkowski, W.; Broniowski, W.
      Melting of the quark condensate in the NJL model with meson loops.
      \emph{Phys.\ Lett.\ B} {\bf 1996}, \emph{386}, 62.
      % doi:10.1016/0370-2693(96)00935-5
      %[hep-ph/9605315].
      %%CITATION = doi:10.1016/0370-2693(96)00935-5;%%
      %29 citations counted in INSPIRE as of 23 Dec 2019
      
      %\cite{Bali:2011qj}
      \bibitem{Bali:2011qj}
      Bali, G.S.; Bruckmann, F.; Endrodi, G.; Fodor, Z.; Katz, S.D.; Krieg, S.; Schafer, A.; Szabo, K.K.
      The QCD phase diagram for external magnetic fields.
      \emph{ J. High Energy Phys.} {\bf2012}, \emph{ 1202}, 044
      %doi:10.1007/JHEP02(2012)044
      %[arXiv:1111.4956 [hep-lat]].
      %%CITATION = doi:10.1007/JHEP02(2012)044;%%
      %457 citations counted in INSPIRE as of 23 Dec 2019
      
        %\cite{Ferreira:2015jrm}
        %\cite{Endrodi:2019whh}
        \bibitem{Endrodi}
        Endrődi, G.; Markó,
        "Magnetized baryons and the QCD phase diagram: NJL model meets the lattice."
        \emph{ J. High Energy Phys.} {\bf 2019}, \emph{1908}, 036.%There are more than one Ref. in the same Reference, please check if the highlight part can be deleted.
        %doi:10.1007/JHEP08(2019)036
        %[arXiv:1905.02103 [hep-lat]];
        %%CITATION = doi:10.1007/JHEP08(2019)036;%%
        %2 citations counted in INSPIRE as of 24 Dec 2019
        
        \bibitem{Ferreira}
       Ferreira, M.R.B.
        	"QCD Phase Diagram Under an External Magnetic Field."
        	Ph.D. Thesis.
        %https://estudogeral.sib.uc.pt/bitstream/10316/29041/1/QCD%20phase%20diagram%20under%20an%20external%20magnetic%20field.pdf
        %%CITATION = INSPIRE-1501993;%%
        
        %\cite{Ferreira:2014kpa}
        \bibitem{Ferreira:2014kpa}
       Ferreira, M.; Costa, P.; Lourenço, O.; Frederico, T.; Providência, C.
        	"Inverse magnetic catalysis in the (2+1)-flavor Nambu-Jona-Lasinio and Polyakov-Nambu-Jona-Lasinio models."
        	\emph{ Phys.\ Rev.\ D} {\bf 2014}, \emph{89}, 116011.
        %  doi:10.1103/PhysRevD.89.116011
        % [arXiv:1404.5577 [hep-ph]].
        %%CITATION = doi:10.1103/PhysRevD.89.116011;%%
        %115 citations counted in INSPIRE as of 24 Dec 2019
        
          %\cite{Mao:2016fha}
          \bibitem{Mao:2016fha}
          Mao, S.
          	Inverse magnetic catalysis in Nambu--Jona-Lasinio model beyond mean field.
          	\emph{Phys.\ Lett.\ B} {\bf2016}, \emph{758}, 195.
          %doi:10.1016/j.physletb.2016.05.018
          % [arXiv:1602.06503 [hep-ph]].
          %%CITATION = doi:10.1016/j.physletb.2016.05.018;%%
          %33 citations counted in INSPIRE as of 23 Dec 2019
          
          
  %\cite{Winstel:2019zfn}
\bibitem{Winstel:2019zfn}
  Winstel, M.; Stoll, J.; Wagner, M.
  "Lattice investigation of an inhomogeneous phase of the 2+1-dimensional Gross-Neveu model in the limit of infinitely many flavors."
  \emph{arXiv} {\bf2019}, arXiv:1909.00064.\\
  %%CITATION = ARXIV:1909.00064;%%
  %1 citations counted in INSPIRE as of 04 Dec 2019
  Pannullo,  L.; Lenz, J.; Wagner,  M.;  Wellegehausen, B.;  Wipf, A.
 "Inhomogeneous phases in the 1+1 dimensional Gross-Neveu model at finite number of fermion flavors." arXiv:1902.11066;\\
 "Lattice investigation of the phase diagram of the 1+1 dimensional Gross-Neveu model at finite number of fermion flavors." arXiv:1909.11513;\\
  %\cite{Pannullo:2019prx}
%\bibitem{Pannullo:2019prx}
 % L.~Pannullo, J.~Lenz, M.~Wagner, B.~Wellegehausen and A.~Wipf,
  "Lattice investigation of the phase diagram of the 1+1 dimensional Gross-Neveu model at finite number of fermion flavors."
  arXiv:1909.11513;
  %%CITATION = ARXIV:1909.11513;%%
  %1 citations counted in INSPIRE as of 04 Dec 2019
 \bibitem{Feinberg}
   Feinberg, J.; Hillel, S.
  "Stable fermion bag solitons in the massive Gross-Neveu model: Inverse scattering analysis."
  \emph{Phys.\ Rev.\ D} {\bf 2005}, \emph{72}, 105009.\\
  %\cite{Vitale:1998wm}
%\bibitem{Vitale:1998wm}
  Vitale, P.
  "Temperature induced phase transitions in four fermion models in curved space-time."
  \emph{Nucl.\ Phys.\ B} {\bf 1999}, \emph{551}, 490.
  %doi:10.1016/S0550-3213(99)00212-6
  %[hep-th/9812076].
  %%CITATION = doi:10.1016/S0550-3213(99)00212-6;%%
  %13 citations counted in INSPIRE as of 04 Oct 2019
  
    %\cite{Gross:1974jv}
\bibitem{Gross:1974jv}
  Gross, D.~J.; Neveu, A.
  Dynamical Symmetry Breaking in Asymptotically Free Field Theories.
  \emph{ Phys.\ Rev.\ D} {\bf 1974}, \emph{10}, 3235.
  %doi:10.1103/PhysRevD.10.3235
  %%CITATION = doi:10.1103/PhysRevD.10.3235;%%
  %1684 citations counted in INSPIRE as of 04 Dec 2019
  
  %\cite{Schnetz:2005ih}
\bibitem{Schnetz:2005ih}
  Schnetz, O.; Thies, M.; Urlichs, K.
  Full phase diagram of the massive Gross-Neveu model.
  \emph{Ann. Phys.}  {\bf 2006}, \emph{321}, 2604.
 % doi:10.1016/j.aop.2005.12.007
 % [hep-th/0511206].
  %%CITATION = doi:10.1016/j.aop.2005.12.007;%%
  %67 citations counted in INSPIRE as of 03 Oct 2019
    %\cite{Thies:2005wv}
    
  %\cite{Caldas:2008zz}
\bibitem{caldas}
  Caldas, H.; Kneur, J.-L.; Pinto, M.~B.; Ramos, R.~O.
  Critical dopant concentration in polyacetylene and phase diagram from a continuous four-Fermi model.
  \emph{Phys.\ Rev.\ B} {\bf 2008}, \emph{77}, 205109.
  %doi:10.1103/PhysRevB.77.205109
  %[arXiv:0804.2675 [cond-mat.soft]].
  %%CITATION = doi:10.1103/PhysRevB.77.205109;%%
  %23 citations counted in INSPIRE as of 04 Dec 2019
  
    
\bibitem{Thies:2005wv}
  Thies, M.; Urlichs, K.
  From non-degenerate conducting polymers to dense matter in the massive Gross-Neveu model.
  \emph{ Phys.\ Rev.\ D} {\bf 2005}, \emph{72}, 105008.
  %doi:10.1103/PhysRevD.72.105008
 % %[hep-th/0505024].
  %%CITATION = doi:10.1103/PhysRevD.72.105008;%%
  %23 citations counted in INSPIRE as of 04 Oct 2019
  
  \bibitem{MertschingFischbeck}
 Mertsching, J.; Fischbeck, H.J.
 The Incommensurate Peierls Phase of the Quasi‐One‐Dimensional Fröhlich Model with a Nearly Half‐Filled Band. 
 \emph{Phys. Stat. Sol. B} {\bf 1981}, \emph{103}, 783.

\bibitem{Machida}
 Machida, K.; Nakanishi, H.
 Superconductivity under a ferromagnetic molecular field.
 \emph{Phys. Rev. B} {\bf 1984}, \emph{30}, 122.
  
  %\cite{Klimenko:2012qi}
\bibitem{pl}
  Caldas, H.; Ramos, R.O.
  "Magnetization of planar four-fermion systems."
  \emph{Phys.\ Rev.\ B} {\bf 2009}, \emph{80}, 115428.%There are more than one Ref. in the same Reference, please check if the highlight part can be deleted.
  
  \bibitem{Klimenko:2012qi}
  Klimenko, K.G.; Zhokhov, R.N.; Zhukovsky, V.C.
  "Superconductivity phenomenon induced by external in-plane magnetic field in (2+1)-dimensional Gross-Neveu type model."
  \emph{Mod.\ Phys.\ Lett.\ A} {\bf 2013}, \emph{28}, 1350096.
%  doi:10.1142/S021773231350096X
 % [arXiv:1211.0148 [hep-th]].
  %%CITATION = doi:10.1142/S021773231350096X;%%
  %9 citations counted in INSPIRE as of 31 Dec 2019
  
  \bibitem{Klimenko:2012tk}
%\bibitem{Klimenko:2012tk}
  K.~G.~Klimenko, R.~N.~Zhokhov and V.~C.~Zhukovsky,
 "Superconducting phase transitions induced by chemical potential in (2+1)-dimensional four-fermion quantum field theory."
  \emph{Phys.\ Rev.\ D} {\bf2012}, \emph{ 86}, 105010.
%  doi:10.1103/PhysRevD.86.105010
 % [arXiv:1210.7934 [hep-th]].
  %%CITATION = doi:10.1103/PhysRevD.86.105010;%%
  %12 citations counted in INSPIRE as of 31 Dec 2019
  
  \bibitem{Klimenko:2013gua}
%\bibitem{Klimenko:2013gua}
  Klimenko, K.G.; Zhokhov, R.N.
  "Magnetic catalysis effect in the (2+1)-dimensional Gross-Neveu model with Zeeman interaction."
  \emph{Phys.\ Rev.\ D} {\bf 2013}, \emph{88}, 105015.
  %doi:10.1103/PhysRevD.88.105015
  %[arXiv:1307.7265 [hep-ph]].
  %%CITATION = doi:10.1103/PhysRevD.88.105015;%%
  %9 citations counted in INSPIRE as of 31 Dec 2019
  
\bibitem{Lin}
Lin,  H.-H.; Balents, L.; Fisher, M.P.A.
"Exact SO(8) Symmetry in the Weakly-Interacting Two-Leg Ladder."
\emph{Phys. Rev. B}  {\bf 1998}, \emph{58}, 1794.%There are more than one Ref. in the same Reference, please check if the highlight part can be deleted.
  %arXiv: 9801285 [cond-mat.str-el];
  
  \bibitem{Kalinkin}
 Kalinkin, A.N.; Shorikov, V. M. 'Phase Transitions in Four-Fermion Models."
 \emph{Inorg. Mater.} {\bf 2003}, \emph{39}, 765.

 \bibitem{kolmakov}
  Zhokhov, R.N.; Zhukovsky, V.C.; Kolmakov, P.B.
  "The Zeeman effect in a modified Gross—Neveu model in (2 + 1)-dimensional space—time with compactification."
  \emph{Moscow Univ.\ Phys.\ Bull.}  {\bf 2015}, \emph{70}, 226.
 % doi:10.3103/S0027134915040165

 %\cite{Thies:2003zr}
\bibitem{Thies:2003zr}
  Thies, M.
  Duality between quark quark and quark anti-quark pairing in 1+1 dimensional large N models.
  \emph{Phys.\ Rev.\ D} {\bf 2003}, \emph{68}, 047703.
  %doi:10.1103/Ph{\bf ysRevD.68.047703
 % [hep-th/0303026];
  %%CITATION = doi:10.1103/PhysRevD.68.047703;%%
  %19 citations counted in INSPIRE as of 04 Dec 2019
  
  %\cite{Basar:2009fg}
\bibitem{Basar:2009fg}
  Basar, G.; Dunne, G.V.; Thies, M.
  "Inhomogeneous Condensates in the Thermodynamics of the Chiral NJL(2) model."
  \emph{ Phys.\ Rev.\ D} {\bf 2009}, \emph{79}, 105012.
  %doi:10.1103/PhysRevD.79.105012
  %[arXiv:0903.1868 [hep-th]].
  %%CITATION = doi:10.1103/PhysRevD.79.105012;%%
  %122 citations counted in INSPIRE as of 04 Dec 2019
  
    %\cite{Thies:2019ejd}
\bibitem{Thies:2019ejd}
  Thies, M.
  Phase structure of the 1+1 dimensional Nambu--Jona-Lasinio model with isospin.
  \emph{arXiv} {\bf 2019}, arXiv:1911.11439.
  %%CITATION = ARXIV:1911.11439;%%
  %\cite{Thies:2016dam}
%\bibitem{Thies:2016dam}
%  M.~Thies,
  %``Solving the U(2)L X U(2)R symmetric Nambu--Jona-Lasinio model in 1+1 dimensions.
  %arXiv:1603.06218 [hep-th];
  %%CITATION = ARXIV:1603.06218;%%
  %3 citations counted in INSPIRE as of 04 Dec 2019
  %\cite{Thies:2019ejd}
%\bibitem{Thies:2019ejd}
 % M.~Thies,
  %``Phase structure of the 1+1 dimensional Nambu--Jona-Lasinio model with isospin.
 % arXiv:1911.11439 [hep-th].
  %%CITATION = ARXIV:1911.11439;%%
  
    \bibitem{kkzz}
  Khunjua, T.G.; Klimenko, K.G.; Zhokhov, R.N.; Zhukovsky, V.C.
  "Inhomogeneous charged pion condensation in chiral asymmetric dense quark matter in the framework of NJL$_2$ model."
  \emph{Phys.\ Rev.\ D} {\bf 2017}, \emph{95}, 105010.%There are more than one Ref. in the same Reference, please check if the highlight part can be deleted.  
  
  % \bibitem{kkzzp}
   % Khunjua, T.G.; Klimenko, K.G.; Zhokhov, R.N.; Zhukovsky, V.C.
   % "Duality and Charged Pion Condensation in Chirally Asymmetric Dense Quark Matter in the Framework of an NJL2 Model"
  %\emph{Int.\ J.\ Mod.\ Phys.\ Conf.\ Ser.}  {\bf 2018}, \emph{47}, 1860093.\\
   %%CITATION = doi:10.1103/PhysRevD.95.105010;%%
     %[arXiv:1808.05162 [hep-ph]].
     
\bibitem{Khunjua:2019ini}
 Khunjua, T.G.; Klimenko, K.G.; Zhokhov, R.N.
  "Charged pion condensation and duality in dense and hot chirally and isospin asymmetric quark matter in the framework of the NJL$_2$ model."
  \emph{Phys.\ Rev.\ D} {\bf 2019}, \emph{100}, 034009.
  %doi:101103/PhysRevD.100.034009
  %[arXiv:1907.04151 [hep-ph]];
  %%CITATION = doi:10.1103/PhysRevD.100.034009;%%
  %1 citations counted in INSPIRE as of 02 Dec 2019
  
  \bibitem{Ebert:2016hkd}
 Ebert,  D.; Khunjua, T.G.; Klimenko, K.G. 
  "Duality between chiral symmetry breaking and charged pion condensation at large $N_c$: Consideration of an NJL$_2$ model with baryon, isospin, and chiral isospin chemical potentials."
  \emph{Phys.\ Rev.\ D} {\bf 2016}, \emph{94},  116016.
  %doi:10.1103/PhysRevD.94.116016
  %[arXiv:1608.07688 [hep-ph]];
  %%CITATION = doi:10.1103/PhysRevD.94.116016;%%
  %15 citations counted in INSPIRE as of 02 Dec 2019
  
  %\cite{Khunjua:2018vyb}
\bibitem{Khunjua:2018vyb}
  T.~G.~Khunjua, V.~C.~Zhukovsky, K.~G.~Klimenko and R.~N.~Zhokhov,
  ``Duality and Charged Pion Condensation in Chirally Asymmetric Dense Quark Matter in the Framework of an NJL2 Model.
  Int.\ J.\ Mod.\ Phys.\ Conf.\ Ser.\  {\bf 47} (2018) 1860093
  %doi:10.1142/S2010194518600935
  %[arXiv:1808.05162 [hep-ph]].
  %%CITATION = doi:10.1142/S2010194518600935;%%
  %3 citations counted in INSPIRE as of 04 Dec 2019
  
    %\cite{Khunjua:2019nnv}
\bibitem{Symmetry}
 Khunjua, T.G.; Klimenko, K.G.; Zhokhov, R.N.
  Charged Pion Condensation in Dense Quark Matter: Nambu--Jona-Lasinio Model Study.
  \emph{Symmetry}  {\bf 2019}, \emph{11},  778.
 %doi:10.3390/sym11060778
  %%CITATION = doi:10.3390/sym11060778;%%
  %2 citations counted in INSPIRE as of 06 Dec 2019
  
  %\cite{Fukushima:2008xe}
\bibitem{Fukushima:2008xe}
Fukushima, K.; Kharzeev, D.E.; Warringa, H.J.
  The Chiral Magnetic Effect.
  \emph{Phys.\ Rev.\ D} {\bf 2008}, \emph{78}, 074033.
  %doi:10.1103/PhysRevD.78.074033
  %[arXiv:0808.3382 [hep-ph]].
  %%CITATION = doi:10.1103/PhysRevD.78.074033;%%
  %1250 citations counted in INSPIRE as of 06 Dec 2019
  
   %\cite{Cherman:2010jj}
\bibitem{Cherman:2010jj}
 Cherman,  A.; Hanada,  M.; Robles-Llana, D.
 "Orbifold equivalence and the sign problem at finite baryon density."
  \emph{Phys.\ Rev.\ Lett.} {\bf 2011}, \emph{106}, 091603.\\%There are more than one Ref. in the same Reference, please check if the highlight part can be deleted.
  %doi:10.1103/PhysRevLett.106.091603
 % [arXiv:1009.1623 [hep-th]];
  %%CITATION = doi:10.1103/PhysRevLett.106.091603;%%
  %42 citations counted in INSPIRE as of 29 Oct 2019
 Cherman, A.; Tiburzi, B.C.
  "The Fermion Sign Problem at Finite Density, and Large Nc Orbifold Equivalence."
  \emph{arXiv} {\bf 2011}, arXiv:1109.3093.
  %%CITATION = ARXIV:1109.3093;%%
  
 % \bibitem{Hanada:2011jb}
 %\hl{ Hanada, M.; Yamamoto, N.} %It seems that ref 35 and ref 37 are the same, please confirm and revise.
 %Universality of phase diagrams in QCD and QCD-like theories.
 % \emph{PoS LATTICE} {\bf 2011}, \emph{\hl{221}}.%Please check if it has the page number and volume number, or doi.
   %https://doi.org/10.22323/1.139.0221
   %https://pos.sissa.it/139/221
 % [arXiv:1111.3391 [hep-lat]].
  %%CITATION = ARXIV:1111.3391;%%
  
  \bibitem{increase}
Hanada, M.; Yamamoto, N.
"Universality of phase diagrams in QCD and QCD-like theories."
%There are more than one Ref. in the same Reference, please check if the highlight part can be deleted.
%\cite{Hanada:2011jb}
%\bibitem{Hanada:2011jb}
%  M.~Hanada and N.~Yamamoto,
  %``Universality of phase diagrams in QCD and QCD-like theories,''
  PoS LATTICE {\bf 2011} (2011) 221.%\emph{arXiv} {\bf2011}, arXiv:1111.3391.
 % doi:10.22323/1.139.0221
 % [arXiv:1111.3391 [hep-lat]].
  %%CITATION = doi:10.22323/1.139.0221;%%
  %19 citations counted in INSPIRE as of 16 Jan 2020
  
  %\cite{Cherman:2011mh}
\bibitem{Cherman:2011mh}
  Cherman, A.; Tiburzi, B.C.
  "Orbifold equivalence for finite density QCD and effective field theory."
  \emph{ J. High Energy Phys.} {\bf 2011}, \emph{1106}, 034.
 % doi:10.1007/JHEP06(2011)034
 % [arXiv:1103.1639 [hep-th]].
  %%CITATION = doi:10.1007/JHEP06(2011)034;%%
  %26 citations counted in INSPIRE as of 29 Oct 2019

\bibitem{increase11}
Andrianov, A.A.; Espriu,  D.; Planells, X.
"Chemical potentials and parity breaking: the Nambu–Jona-Lasinio model."
\emph{Eur. Phys. J. C}  {\bf 2014}, \emph{74}, 2776.

\bibitem{increase111}
Wang, Y.-L.; Cui, Z.-F.; Zong, H.-S.
"Effect of the chiral chemical potential on the position of the critical endpoint."
\emph{Phys. Rev. D} {\bf 2015}, \emph{91}, 034017.

\bibitem{increase1}
Xu, S.-S.; Cui, Z.-F.; Wang, B.; Shi, Y.-M.; Yang, Y.-C.;  Zong, H.-S.
"Chiral phase transition with a chiral chemical potential in the framework of Dyson-Schwinger equations."
\emph{Phys. Rev. D} {\bf 2015}, \emph{91}, 056003.

\bibitem{increase2}
Ruggieri, M.; Peng, G.X.
%\emph{arXiv} {\bf2016}, arXiv:1602.05250.}
%\cite{Ruggieri:2016ejz}
%\bibitem{Ruggieri:2016ejz}
%  M.~Ruggieri and G.~X.~Peng,
  "Critical Temperature of Chiral Symmetry Restoration for Quark Matter with a Chiral Chemical Potential.''
  J.\ Phys.\ G {\bf 43} (2016) no.12,  125101
 % doi:10.1088/0954-3899/43/12/125101
%  [arXiv:1602.05250 [hep-ph]].
  %%CITATION = doi:10.1088/0954-3899/43/12/125101;%%
  %20 citations counted in INSPIRE as of 16 Jan 2020
  
\bibitem{increase3}
Frasca, M.
"Nonlocal Nambu-Jona-Lasinio model and chiral chemical potential."
\emph{Eur. Phys. J. C} {\bf 2018}, \emph{78}, 790; %\emph{arXiv} {\bf2018}, arXiv:1602.04654.

\bibitem{increase4}
Ruggieri, M.; Peng, G.X.
"Critical Temperature of Chiral Symmetry Restoration for Quark Matter with a Chiral Chemical Potential."
\emph{arXiv} {\bf2016}, arXiv:1602.03651.

\bibitem{increase5}
Farias, R.L.S.; Duarte, D.C.; Krein, G.; Ramos, R.O.
"Thermodynamics of quark matter with a chiral imbalance."
%\emph{arXiv} {\bf2016}, arXiv:1604.04518.
%\cite{Farias:2016let}
%\bibitem{Farias:2016let}
 % R.~L.~S.~Farias, D.~C.~Duarte, G.~Krein and R.~O.~Ramos,
  %``Thermodynamics of quark matter with a chiral imbalance,''
  Phys.\ Rev.\ D {\bf 94} (2016) no.7,  074011;
  %doi:10.1103/PhysRevD.94.074011
  [arXiv:1604.04518 [hep-ph]].
  %%CITATION = doi:10.1103/PhysRevD.94.074011;%%
  %19 citations counted in INSPIRE as of 16 Jan 2020
  
\bibitem{increase55}
Xu, S.S.; Cui, Z.F.; Wang, B.; Shi, Y.M.; Yang, Y.C.; Zong, H.S.
  "Chiral phase transition with a chiral chemical potential in the framework of Dyson-Schwinger equations."
  \emph{ Phys.\ Rev.\ D} {\bf 2015}, \emph{91}, 056003.

\bibitem{decrease}
Fukushima, K.; Ruggieri, M.; Gatto, R.
"Chiral magnetic effect in the Polyakov--Nambu--Jona-Lasinio model."
\emph{Phys. Rev. D} {\bf 2010}, \emph{81}, 114031.%There are more than one Ref. in the same Reference, please check if the highlight part can be deleted.

\bibitem{decrease1}
Chernodub, M.N.; Nedelin, A.S.
"Phase diagram of chirally imbalanced QCD matter."
\emph{Phys. Rev. D} {\bf 2011}, \emph{83}, 105008.

\bibitem{decrease2}
Gatto, R.; Ruggieri,  M.
"Hot quark matter with an axial chemical potential."
\emph{Phys.Rev. D} {\bf 2012}, \emph{85}, 054013.

\bibitem{decrease3}
Ruggieri, M.
  ""The Critical End Point of Quantum Chromodynamics Detected by Chirally Imbalanced Quark Matter.""
  \emph{Phys.\ Rev.\ D} {\bf 2011}, \emph{84}, 014011.

\bibitem{decrease4}
Ruggieri, M.; Peng, G.X.
  "Critical Temperature of Chiral Symmetry Restoration for Quark Matter with a Chiral Chemical Potential."
 \emph{arXiv} {\bf2016}, arXiv:1602.03651.

\bibitem{decrease5}
Chao, J.; Chu, P.; Huang, M.
"Inverse magnetic catalysis induced by sphalerons."
\emph{Phys. Rev. D} {\bf 2013}, \emph{88}, 054009.

\bibitem{decrease6}
Yu, L.; Liu, H.; Huang, M.
"Spontaneous generation of local CP violation and inverse magnetic catalysis."
\emph{Phys. Rev. D} {\bf 2014}, \emph{90}, 074009.

\bibitem{decrease7}
Yu, L.; Liu, H.; Huang, M.
"Effect of the chiral chemical potential on the chiral phase transition in the NJL model with different regularization schemes."
\emph{Phys. Rev. D} {\bf 2016}, \emph{94}, 014026.

\bibitem{decrease8}
%, arXiv:1511.03073;
Cui, F.; Cloet, I.C.; Lu, Y.; Roberts, C.D.; Schmidt, S.M.; Xu, S.S.; Zong, H.S.
  "Critical endpoint in the presence of a chiral chemical potential."
  \emph{ Phys.\ Rev.\ D} {\bf 2016}, \emph{94}, 071503.
%  doi:10.1103/PhysRevD.94.071503
 % [arXiv:1604.08454 [nucl-th]].
  %%CITATION = doi:10.1103/PhysRevD.94.071503;%%
%\cite{Hanada:2011ju}

  \bibitem{Son} 
  Son,D.T.; Stephanov, M.A.
  "QCD at finite isospin density."
  \emph{Phys.\ Rev.\ Lett.}  {\bf 2001}, \emph{86}, 592.\\
 % doi:10.1103/PhysRevLett.86.592
  %[hep-ph/0005225].
  %%CITATION = doi:10.1103/PhysRevLett.86.592;%%
  "QCD at finite isospin density: from pion to quark-antiquark condensation"
  \emph{Phys.\ Atom.\ Nucl.} {\bf 2001}, \emph{64}, 834.;
  \emph{Yad.\ Fiz.} {\bf 2001}, \emph{64}, 899.
 % doi:10.1134/1.1378872
  %[hep-ph/0011365].
  
  \bibitem{Loewe} 
Loewe,M.; Villavicencio, C.
  "Thermal pions at finite isospin chemical potential."
  \emph{ Phys.\ Rev.\ D} {\bf 2003}, \emph{67}, 074034.
  %doi:10.1103/PhysRevD.67.074034
  %[hep-ph/0212275].
  %%CITATION = doi:10.1103/PhysRevD.67.074034;%%

%\cite{Adhikari:2019mlf}
\bibitem{Adhikari:2019mlf}
 Adhikari, P.; Andersen, J.O.
  "Pion and kaon condensation at zero temperature in three-flavor $\chi$PT at nonzero isospin and strange chemical potentials at next-to-leading order."
  \emph{arXiv} {\bf2019}, arXiv:1909.10575.\\%There are more than one Ref. in the same Reference, please check if the highlight part can be deleted.
  %%CITATION = ARXIV:1909.10575;%%
  %2 citations counted in INSPIRE as of 06 Dec 2019
  %\cite{Adhikari:2019zaj}
%\bibitem{Adhikari:2019zaj}
 % P.~Adhikari and J.~O.~Andersen,
  %``QCD at finite isospin density: chiral perturbation theory confronts lattice data.
  %arXiv:1909.01131 [hep-ph];
  %%CITATION = ARXIV:1909.01131;%%
  %3 citations counted in INSPIRE as of 06 Dec 2019
  %\cite{Adhikari:2019mdk}
%\bibitem{Adhikari:2019mdk}
 Adhikari, P.; Andersen, J.O.; Kneschke, P.
  "Two-flavor chiral perturbation theory at nonzero isospin: Pion condensation at zero temperature."
  \emph{Eur.\ Phys.\ J.\ C}  {\bf 2019}, \emph{79},  874.
 % doi:10.1140/epjc/s10052-019-7381-4
  %[arXiv:1904.03887 [hep-ph]].
  %%CITATION = doi:10.1140/epjc/s10052-019-7381-4;%%
  %6 citations counted in INSPIRE as of 06 Dec 2019

\bibitem{buballa}
Buballa, M.; Carignano, S.
  Inhomogeneous chiral condensates.
  \emph{Prog.\ Part.\ Nucl.\ Phys.} {\bf 2015}, \emph{81}, 39.
 % doi:10.1016/j.ppnp.2014.11.001
 % [arXiv:1406.1367 [hep-ph]].
  %%CITATION = doi:10.1016/j.ppnp.2014.11.001;%%
  
  \bibitem{Heinz:2014pkl}
 Heinz, A.
 QCD under Extreme Conditions: Inhomogeneous Condensation.
%Dissertation zur Erlangung des Doktorgrades der Naturwissenschaften,
Ph.D Thesis, Frankfurt University, Frankfurt am Main, September 2014.
%Please add the collabary of the thesis.
  %%CITATION = INSPIRE-1411975;%%
  %https://inis.iaea.org/search/search.aspx?orig_q=RN:47053579
  
  \bibitem{Nickel:2009wj}
  Nickel, D.
  Inhomogeneous phases in the Nambu-Jona-Lasino and quark-meson model.
  \emph{Phys.\ Rev.\ D} {\bf 2009}, \emph{80}, 074025.
    %%CITATION = doi:10.1103/PhysRevD.80.074025;%%
    
    \bibitem{Nowakowski:2015ksa}
 Nowakowski,  D.; Buballa, M.; Carignano,  S.; Wambach, J.
  Inhomogeneous chiral symmetry breaking phases in isospin-asymmetric matter.
  \emph{arXiv} {\bf2015}, arXiv:1506.04260.
  %%CITATION = ARXIV:1506.04260;%%

\bibitem{Nowakowski}
Nowakowski, D.
Inhomogeneous Chiral Symmetry Breaking in Isospin-Asymmetric Strong-Interaction Matter. 
Ph.D. Thesis, Technische Univ. Darmstadt, Darmstadt, Germany, 2017.

 \bibitem{Andersen:2018osr}
  Andersen, J.~O.; Kneschke, P.
  Chiral density wave versus pion condensation at finite density and zero temperature.
  \emph{ Phys.\ Rev.\ D} {\bf 2018}, \emph{97}, 076005.
  %doi:10.1103/PhysRevD.97.076005
  %[arXiv:1802.01832 [hep-ph]].
  %%CITATION = doi:10.1103/PhysRevD.97.076005;%%

\bibitem{he}
Mu, C.F.  He,  L. Y.; Liu, Y.X.
Evaluating the phase diagram at finite isospin and baryon chemical potentials in the Nambu--Jona-Lasinio model.
\emph{ Phys.\ Rev.\ D} {\bf 2010} \emph{82}, 056006.
    %\cite{Khunjua:2019lbv}
\bibitem{Khunjua:2019lbv}
  Khunjua, T.G.; Klimenko, K.G.; Zhokhov, R.N.
  Dualities and inhomogeneous phases in dense quark matter with chiral and isospin imbalances in the framework of effective model.
  \emph{ J. High Energy Phys.} {\bf 2019}, \emph{1906}, 006.
  
  %doi:10.1007/JHEP06(2019)006
  %[arXiv:1901.02855 [hep-ph]].
  %%CITATION = doi:10.1007/JHEP06(2019)006;%%
  %3 citations counted in INSPIRE as of 04 Dec 2019
  
    %\cite{Anglani:2006br}
\bibitem{supercond}
    Shovkovy, I.A.
  Two lectures on color superconductivity.
  \emph{Found.\ Phys.}  {\bf 2005}, \emph{35}, 1309.
 % doi:10.1007/s10701-005-6440-x
%  [nucl-th/0410091].
  
  %\cite{Shovkovy:2004me}
\bibitem{inhomsupercond}
  Anglani, R.; Nardulli, G.; Ruggieri, M.; Mannarelli, M.
  "Neutrino emission from compact stars and inhomogeneous color superconductivity."
  \emph{ Phys.\ Rev.\ D} {\bf 2006}, \emph{74}, 074005.
%  doi:10.1103/PhysRevD.74.074005
  %[hep-ph/0607341];
  %%CITATION = doi:10.1103/PhysRevD.74.074005;%%
  %45 citations counted in INSPIRE as of 30 Dec 2019
  
   \bibitem{Anglani:2013gfu}
%\bibitem{Anglani:2013gfu}
 Anglani, R.; Casalbuoni, R.; Ciminale, M.; Ippolito, N.; Gatto, R.; Mannarelli, M.; Ruggieri, M.
  "Crystalline color superconductors."
 \emph{Rev.\ Mod.\ Phys.}  {\bf 2014}, \emph{86}, 509.
  %doi:10.1103/RevModPhys.86.509
 % [arXiv:1302.4264 [hep-ph]].
  %%CITATION = doi:10.1103/RevModPhys.86.509;%%
  %96 citations counted in INSPIRE as of 30 Dec 2019
  
      %\cite{Khunjua:2018ant}
\bibitem{Khunjua:2018ant}
 Khunjua, T.G.; Klimenko, K.G.; Zhokhov, R.N.
 Dense quark matter with chiral and isospin imbalance: NJL-model consideration.
 \emph{EPJ Web Conf.} {\bf 2018}, \emph{191}, 05015.%There are more than one Ref. in the same Reference, please check if the highlight part can be deleted.
  %doi:10.1051/epjconf/201819105015
  %[arXiv:1901.03049 [hep-ph]];
  %%CITATION = doi:10.1051/epjconf/201819105015;%%
  %3 citations counted in INSPIRE as of 04 Dec 2019
  
  \bibitem{Khunjua:2018xly}
  Khunjua, T.G.; Klimenko, K.G.; Zhokhov-Larionov, R.N.  "Affinity of NJL$_2$ and NJL$_4$ model results on duality and pion condensation in chiral asymmetric dense quark matter."
  \emph{EPJ Web Conf.} {\bf 2018}, \emph{191}, 05016.
  %doi:10.1051/epjconf/201819105016
  %[arXiv:1812.01860 [hep-ph]].
  %%CITATION = doi:10.1051/epjconf/201819105016;%%
  %3 citations counted in INSPIRE as of 04 Dec 2019
  
  
  \bibitem{kkz3}
  Khunjua, T.G.; Klimenko, K.G.; Zhokhov, R.N.
  Dualities in dense quark matter with isospin, chiral and chiral isospin imbalance in the framework of the large-N$_{c}$ limit of the NJL$_{4}$ model.
  \emph{ Phys.\ Rev.\ D} {\bf 2018}, \emph{98}, 054030.%There are more than one Ref. in the same Reference, please check if the highlight part can be deleted.
 % doi:10.1103/PhysRevD.98.054030
  %[arXiv:1804.01014 [hep-ph]].
  %%CITATION = doi:10.1103/PhysRevD.98.054030;%%
  
   \bibitem{kkz4}
T.~G.~Khunjua, K.~G.~Klimenko and R.~N.~Zhokhov,
  ``Dense baryon matter with isospin and chiral imbalance in the framework of NJL$_4$ model at large $N_c$: duality between chiral symmetry breaking and charged pion condensation.
  \emph{ Phys.\ Rev.\ D} {\bf 2018}, \emph{97}, 054036.
  %%CITATION = doi:10.1103/PhysRevD.97.054036;%%
 
   %\cite{Khunjua:2019ojm}
\bibitem{Khunjua:2019ojm}
  Khunjua, T.~G. Klimenko, K.~G. and Zhokhov, R.~N.
  "Quark/Hadronic Matter and Dualities of QCD Thermodynamics."
  arXiv:1912.09102 [hep-ph].
  %%CITATION = ARXIV:1912.09102;%%
 
 
   % *****
  
  %\cite{Walters:2004dq}
%\bibitem{Walters:2004dq}
%  Walters, D.N.
%  Cold, dense matter via the lattice NJL model.
%  \emph{arXiv} {\bf2015}, arXiv:hep-lat/0408042.
  %%CITATION = HEP-LAT/0408042;%%
  
  
  %%%%%%%%%%%%%%%%%%%%%%%%%%%%%%%%%%%%%%%%%%%%%%%%%%%%%%%%%%
  
    %\cite{TheLIGOScientific:2017qsa}
\bibitem{TheLIGOScientific:2017qsa}
  Abbott, B.P.; Abbott, R.; Abbott, T.D.; Acernese, F.; Ackley, K.; Adams, C.; Adams, T.; Addesso, P.; Adhikari,~R.X.; Adya, V.B.; et~al.
  GW170817: Observation of Gravitational Waves from a Binary Neutron Star Inspiral.
  \emph{Phys.\ Rev.\ Lett.}  {\bf 2017}, \emph{119},  161101.\\%There are more than one Ref. in the same Reference, please check if the highlight part can be deleted.
%  doi:10.1103/PhysRevLett.119.161101
%  [arXiv:1710.05832 [gr-qc]];
  %%CITATION = doi:10.1103/PhysRevLett.119.161101;%%
  %2756 citations counted in INSPIRE as of 25 Dec 2019
  %\cite{GBM:2017lvd}
%\bibitem{GBM:2017lvd}
  Abbott, B.P.; et al. %[LIGO Scientific and Virgo and Fermi GBM and INTEGRAL and IceCube and IPN and Insight-Hxmt and ANTARES and Swift and Dark Energy Camera GW-EM and DES and DLT40 and GRAWITA and Fermi-LAT and ATCA and ASKAP and OzGrav and DWF (Deeper Wider Faster Program) and AST3 and CAASTRO and VINROUGE and MASTER and J-GEM and GROWTH and JAGWAR and CaltechNRAO and TTU-NRAO and NuSTAR and Pan-STARRS and KU and Nordic Optical Telescope and ePESSTO and GROND and Texas Tech University and TOROS and BOOTES and MWA and CALET and IKI-GW Follow-up and H.E.S.S.; LOFAR and LWA and HAWC and Pierre Auger and ALMA and Pi of Sky and DFN and ATLAS Telescopes and High Time Resolution Universe Survey and RIMAS and RATIR and SKA South Africa/MeerKAT Collaborations and AstroSat Cadmium Zinc Telluride Imager Team and AGILE Team and 1M2H Team and Las Cumbres Observatory Group and MAXI Team and TZAC Consortium and SALT Group and Euro VLBI Team and Chandra Team at McGill University],
  "Multi-messenger Observations of a Binary Neutron Star Merger."
  \emph{Astrophys.\ J.}  {\bf 2017}, \emph{848},  L12.
  %doi:10.3847/2041-8213/aa91c9
 % [arXiv:1710.05833 [astro-ph.HE]].
  %%CITATION = doi:10.3847/2041-8213/aa91c9;%%
  %1132 citations counted in INSPIRE as of 25 Dec 2019
  
  %\cite{Galeazzi:2013mia}
\bibitem{Galeazzi:2013mia}
  Galeazzi, F.; Kastaun, W.; Rezzolla, L.; Font, J.A.
  Implementation of a simplified approach to radiative transfer in general relativity.
  \emph{ Phys.\ Rev.\ D} {\bf 2013}, \emph{88}, 064009.%There are more than one Ref. in the same Reference, please check if the highlight part can be deleted.
 % doi:10.1103/PhysRevD.88.064009
  %[arXiv:1306.4953 [gr-qc]];
  %%CITATION = doi:10.1103/PhysRevD.88.064009;%%
  %65 citations counted in INSPIRE as of 24 Dec 2019
  
  %\cite{Dexheimer:2017nse}
\bibitem{Dexheimer:2017nse}
 Dexheimer, V.
  "Tabulated Neutron Star Equations of State Modeled within the Chiral Mean Field Model."
  Publications of the Astronomical Society of Australia 34 (2017);
 % doi:10.1017/pasa.2017.61
  arXiv:1708.08342.
  %%CITATION = doi:10.1017/pasa.2017.61;%%
  %3 citations counted in INSPIRE as of 24 Dec 2019
  
    %\cite{Alford:2018lhf}
\bibitem{Alford:2018lhf}
  Alford, M.G.; Harris, S.P.
  %``Beta equilibrium in neutron star mergers.
  \emph{Phys.\ Rev.\ C} {\bf2018}, {\bf 98},  065806.
  %doi:10.1103/PhysRevC.98.065806.
%  [arXiv:1803.00662 [nucl-th]].
  %%CITATION = doi:10.1103/PhysRevC.98.065806;%%
  %8 citations counted in INSPIRE as of 24 Dec 2019
  
  %\cite{Fischer:2010wp}
\bibitem{Fischer:2010wp}
  %T.~Fischer {\it et al.},
Fischer, T.; Blaschke, D.; Hempel, M.; Klahn, T.; Lastowiecki, R.; Liebendorfer, M.; Martinez-Pinedo, G.; Pagliara, G.; Sagert, I.; Sandin, F.; et al.
Core-collapse supernova explosions triggered by a quark-hadron phase transition during the early post-bounce phase.
\emph{Astrophys.\ J.\ Suppl.}  {\bf 2011}, \emph{194}, 39.\\
 % doi:10.1088/0067-0049/194/2/39
%  [arXiv:1011.3409 [astro-ph.HE]].
  %%CITATION = doi:10.1088/0067-0049/194/2/39;%%
  %98 citations counted in INSPIRE as of 25 Dec 2019
  %\cite{Fischer:2011zj}
%\bibitem{Fischer:2011zj}
  %T.~Fischer {\it et al.}, 
 % T. Fischer, D. Blaschke, M. Hempel, T. Klahn, R. Lastowiecki, M. Liebendorfer, G. Martinez-Pinedo, G. Pagliara, I. Sagert, F. Sandin, J. Schaffner-Bielich, S. Typel
  "Core collapse supernovae in the QCD phase diagram."
  \emph{Phys.\ Atom.\ Nucl.}  {\bf 2012}, \emph{75}, 613.
 % doi:10.1134/S1063778812050067
 % [arXiv:1103.3004 [astro-ph.HE]].
  %%CITATION = doi:10.1134/S1063778812050067;%%
  %15 citations counted in INSPIRE as of 25 Dec 2019
  
    %\cite{Ohnishi:2011jv}
\bibitem{Ohnishi:2011jv}
  Ohnishi, A.; Ueda, H.; Nakano, T.Z.; Ruggieri, M.; Sumiyoshi, K.
  Possibility of QCD critical point sweep during black hole formation.
  \emph{Phys.\ Lett.\ B} {\bf 2011}, \emph{704}, 284.\\
 % doi:10.1016/j.physletb.2011.09.018
  %[arXiv:1102.3753 [nucl-th]];
  %%CITATION = doi:10.1016/j.physletb.2011.09.018;%%
  %27 citations counted in INSPIRE as of 25 Dec 2019
  %\cite{Sumiyoshi:2008kw}
%\bibitem{Sumiyoshi:2008kw}
 Sumiyoshi, K.; Ishizuka, C.; Ohnishi, A.; Yamada, S.; Suzuki, H.
  "Emergence of hyperons in failed supernovae: trigger of the black hole formation."
  \emph{Astrophys.\ J.} {\bf 2009}, \emph{690}, L43.
%  doi:10.1088/0004-637X/690/1/L43
  %[arXiv:0811.4237 [astro-ph]].
  %%CITATION = doi:10.1088/0004-637X/690/1/L43;%%
  %44 citations counted in INSPIRE as of 25 Dec 2019
  
  %\cite{Sumiyoshi:2006id}
\bibitem{Sumiyoshi:2006id}
  Sumiyoshi, K.; Yamada, S.; Suzuki, H.; Chiba, S.
  Neutrino signals from the formation of black hole: A probe of equation of state of dense matter.
  \emph{Phys.\ Rev.\ Lett.}  {\bf 2006}, \emph{97}, 091101.
  %doi:10.1103/PhysRevLett.97.091101.
  %[astro-ph/0608509].
  %%CITATION = doi:10.1103/PhysRevLett.97.091101;%%
  %80 citations counted in INSPIRE as of 25 Dec 2019
\end{thebibliography}
\end{document}